\documentclass[10pt]{article}

\usepackage{fullpage}
\usepackage{CJKutf8} 
\usepackage[utf8]{inputenc}
\usepackage{setspace}
\usepackage{parskip}
\usepackage{titlesec}
\usepackage[section]{placeins}
\usepackage{xcolor}
\usepackage{breakcites}
\usepackage{lineno}

\usepackage{hyphenat}

\PassOptionsToPackage{hyphens}{url}
\usepackage[colorlinks = true,
            linkcolor = blue,
            urlcolor  = blue,
            citecolor = blue,
            anchorcolor = blue]{hyperref}
\usepackage{etoolbox}
\makeatletter

\usepackage[numbers, sort&compress]{natbib}

\renewenvironment{abstract}
  {{\bfseries\noindent{\abstractname}\par\nobreak}\footnotesize}
  {\bigskip}

\titlespacing{\section}{0pt}{*3}{*1}
\titlespacing{\subsection}{0pt}{*2}{*0.5}
\titlespacing{\subsubsection}{0pt}{*1.5}{0pt}

\usepackage{authblk}

\usepackage{graphicx}
\usepackage[space]{grffile}
\usepackage{latexsym}
\usepackage{textcomp}
\usepackage{longtable}
\usepackage{tabulary}
\usepackage{booktabs,array,multirow}
\usepackage{amsfonts,amsmath,amssymb}
\providecommand\citet{\cite}
\providecommand\citep{\cite}

\newif\iflatexml\latexmlfalse

\AtBeginDocument{\DeclareGraphicsExtensions{.pdf,.PDF,.eps,.EPS,.png,.PNG,.tif,.TIF,.jpg,.JPG,.jpeg,.JPEG}}

\usepackage[utf8]{inputenc}
\usepackage[greek,english]{babel}

\begin{document}
\begin{CJK}{UTF8}{gbsn}

\title{scI$^2$CL: Effectively Integrating Single-cell Multi-omics by Intra- and Inter-omics Contrastive Learning}

\author[1]{Wuchao Liu}%
\author[1]{Han Peng}%
\author[1]{Wengen Li}%
\author[1]{Yichao Zhang}%
\author[1,$\ast$]{Jihong Guan}%
\author[2,$\ast$]{Shuigeng Zhou}%

\affil[1]{School of Computer Science and Technology, Tongji University, Shanghai 201804, China}%
\affil[2]{College of Computer Science and Artificial Intelligence, Fudan University, Shanghai 200438, China}%
\affil[*]{Corresponding author. \href{email:email1}{jhguan@tongji.edu.cn}
and \href{email:email2}{sgzhou@fudan.edu.cn}}

\vspace{-1em}


\begingroup
\let\center\flushleft
\let\endcenter\endflushleft
\maketitle
\endgroup

\selectlanguage{english}
\begin{abstract}
 Single-cell multi-omics data contain huge information of cellular states, and analyzing these data can reveal valuable insights into cellular heterogeneity, diseases, and biological processes. However, as cell differentiation \& development is a continuous and dynamic process, it remains challenging to computationally model and infer cell interaction patterns based on single-cell multi-omics data. This paper presents scI$^2$CL, a new single-cell multi-omics fusion framework based on intra- and inter-omics contrastive learning, to learn comprehensive and discriminative cellular representations from complementary multi-omics data for various downstream tasks. Extensive experiments of four downstream tasks validate the effectiveness of scI$^2$CL and its superiority over existing peers. Concretely, in cell clustering, scI$^2$CL surpasses eight state-of-the-art methods on four widely-used real-world datasets. In cell subtyping, scI$^2$CL effectively distinguishes three latent monocyte cell subpopulations, which are not discovered by existing methods.  
Simultaneously, scI$^2$CL is the only method that correctly constructs the cell developmental trajectory from hematopoietic stem and progenitor cells to Memory B cells.
In addition, scI$^2$CL resolves the misclassification of cell types between two subpopulations of CD4+ T cells, while existing methods fail to precisely distinguish the mixed cells. In summary, scI$^2$CL can accurately characterize cross-omics relationships among cells, thus effectively fuses multi-omics data and learns discriminative cellular representations to support various downstream analysis tasks.
\\%
\end{abstract}%

\sloppy







\section{Introduction}

In the past decade, single-cell sequencing technologies have achieved significant advances, enabling scientists to quantitatively measure and analyze various types of omics data at the single-cell level, providing valuable insights to cellular heterogeneity, biological processes, and human disease~\cite{ref1,ref2,ref3}. Specifically, single-cell RNA-sequencing (scRNA-seq) quantifies the gene expression levels in cells, while single-cell Assay for Transposase Accessible Chromatin sequencing~(scATAC-seq) explores the openness of cis-regulation elements in nearby genes~\cite{scmvae}.
The joint analysis of scRNA-seq and scATAC-seq data can capture complementary information, enabling a more comprehensive characterization of cellular heterogeneity and gene regulatory relationships~\cite{ref4}. And integrated representations derived from multi-omics data drive comprehensive modeling of cellular states and cross-omics heterogeneity, has thereby been proved critical for precisely identifying cell types and functions, finding latent subpopulations, and inferring developmental trajectories.

Although single-cell omics technologies such as SNARE-seq~\cite{chen2019} and 10X Genomics Multiome~\cite{10x} provide opportunities to interrogate cells across multiple omics, the integration of single-cell multi-omics remains significantly challenging due to high sparsity, increasingly large scale~\cite{ref6}, and considerable inter-omics discrepancies~\cite{ref5}.
Deep learning, with its capacity to capture patterns and representations in complex high-dimensional data, has emerged as the mainstream approach for single-cell data analysis~\cite{ref7}. Early works have focused on single-omic data analysis. For example, scVI~\cite{scvi} adapts a variational autoencoder supervised by zero-inflated negative binomial distribution for scRNA-seq analysis. scDSC~\cite{scdsc} unifies an autoencoder and a graph neural network to perform cell clustering with scRNA-seq data. PeakVI~\cite{peakvi} leverages variational inference based deep neural networks to model scATAC-seq data. scBasset~\cite{scbasset} employs a sequence-based convolutional neutral network for scATAC-seq analysis. 
However, these methods are all restricted to single omics analysis, which prevents them from leveraging consistent and complementary information across multi-omics to obtain biologically coherent cellular representations~\cite{multiomics_1,multiomics_2}.

Recently, more and more efforts have endeavored to model and explore cells by exploiting single-cell multi-omics. 
To name a few, scMVAE utilizes probabilistic Gaussian mixture models to explore latent features that characterize multi-omics data~\cite{scmvae}. Wang et al.~\cite{vimcca} introduced a variational-assisted multi-view canonical correlation analysis~(VIMCCA) model based on variational inference and multi-view subspace learning to explore cell identities and functions from paired multi-omics data. Zuo et al.~\cite{dcca} developed the deep cross-omics cycle attention~(DCCA) model, which employs an attention-transfer mechanism to integrate multi-omics featrues extracted by omic-specific variational autoencoders. scMCs~\cite{scmcs} is a single-cell data fusion based clustering method, which incorporates omics-label discriminator, attention mechanism and cross-omic contrastive learning within an autoencoder framework for multi-omics data integration.

Although these methods above have achieved promising results in integrating single-cell multi-omics data, they are limited in at least two aspects. 
On the one hand, most existing methods simply use an encoder-decoder structure to learn features from single-omics count matrices, without sufficiently accounting for potential local dependencies (e.g. the joint regulatory effects among multiple genes in scRNA-seq data). 
On the other hand, current methods predominantly focus on capturing interdependencies between paired omics, overlooking the potential relationships among the more abundant unmatched multi-omics. The high noise and batch effects within single-cell data may obscure the relational patterns between different omics, making it difficult for features learned solely from paired data to effectively model the interdependencies among omics. This hinders the models from capturing both commonality and individuality across different omics, leading to inadequate cellular state representations, ultimately restricts their capacity for various downstream tasks.

To overcome these limitations, this paper presents scI$^2$CL, a novel single-cell multi-omics fusion framework that combines intra- and inter-omics contrastive learning for various downstream single-cell omics data analysis. 
To address the first limitation, we introduce intra-omics contrastive learning to extract both local and global features in single-omics data. To cope with the second limitation, we propose inter-omics contrastive learning that utilizes the unmatched cross-omic relationships to mitigate noise perturbations during data integration. Our approach preserves the individuality of single-omics features while capturing the common dependencies across different omics, ultimately yielding comprehensive and discriminative cellular representations. 

Our extensive experiments of four different downstream tasks show that scI$^2$CL is not only effective, but also superior to existing methods. Concretely, 
in cell clustering, scI$^2$CL outperforms eight state-of-the art methods in terms of both Normalized Mutual Information~(NMI) and Adjusted Rand Index~(ARI) on four real world single-cell multi-omics datasets. In cell subpopulation analysis, scI$^2$CL identifies from the PBMC-10K dataset three CD14 Monocyte subpopulations undetected by existing computational methods. For cell annotation correction, scI$^2$CL precisely recognizes misannotated CD4+ T cell types in the PBMC-3K dataset, which are hard for peer methods to discriminate. Additionly, with the integrated representations learned by scI$^2$CL, we successfully reconstruct the complete hematopoietic trajectory from hematopoietic stem progenitor cells to B cell lineages in the PBMC-10K dataset, but fail when using existing methods. 

In summary, by leveraging the capabilities of intra- and inter-omics contrastive learning, scI$^2$CL effectively discovers intra-omics dependencies while capturing latent biologically relevant patterns across inter-omic samples, thereby achieving effective multi-omic integration and learn representations favorable to precisely delineate cellular heterogeneity. 
scI$^2$CL's powerful multi-omics integration capability provides new opportunities for delineating disease subtypes, elucidating fundamental biological mechanisms, and enabling personalized diagnostic and therapeutic strategies.

\section{Results}
\subsection{Overview of the scI$^2$CL framework}
\begin{figure*}
\centering
\includegraphics[width=0.9\textwidth]{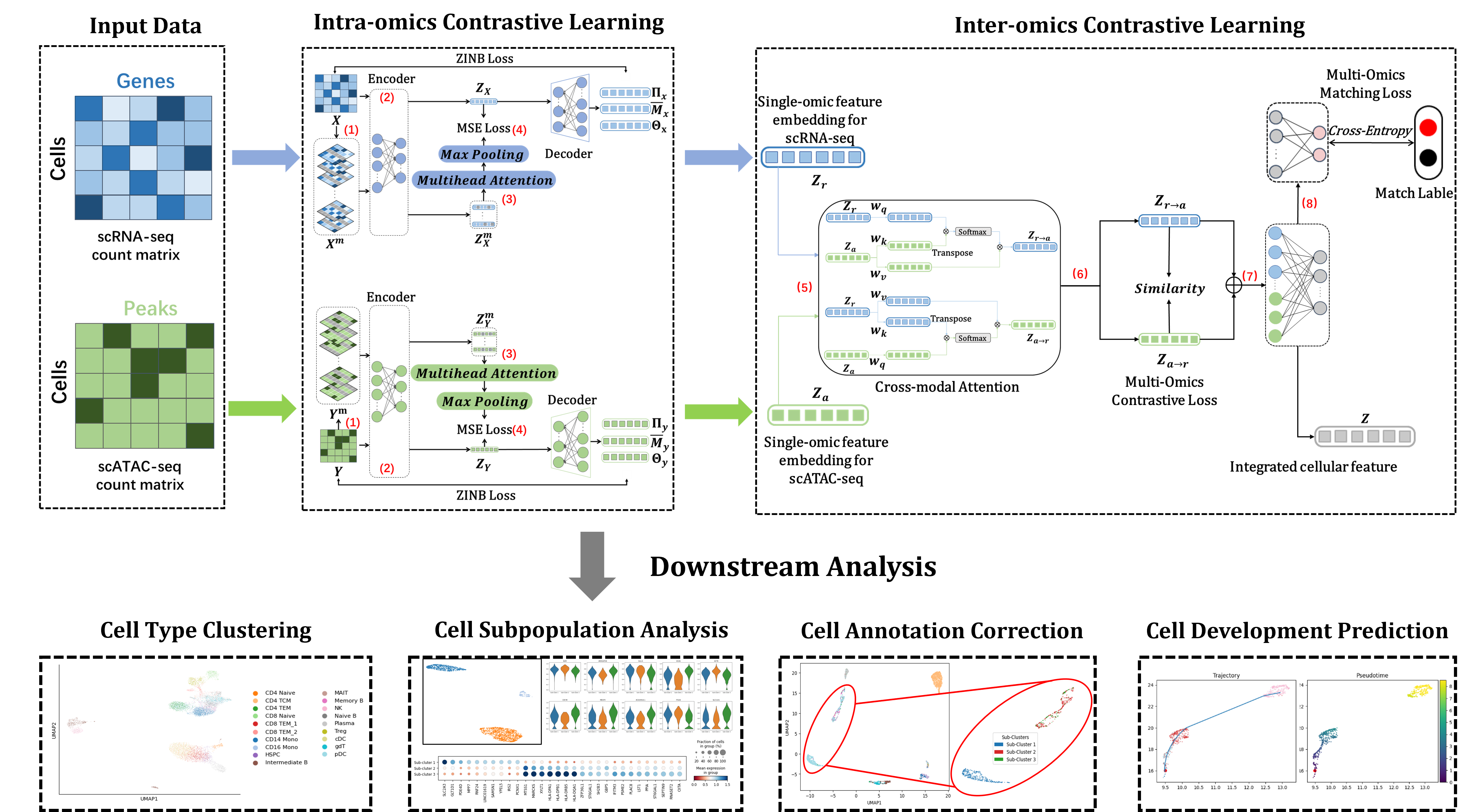}
\caption{The scI$^2$CL framework consists of two major modules: intra-omics contrastive learning (Intra-CL) and inter-omics contrastive learning (Inter-CL). Initially, Intra-CL generates multi-view matrices and extracts representations from the original count matrix and corresponding multi-view matrices using zero-inflated negative binomial distribution-based auto-encoders and multi-head attention. A soft fusion strategy, based on mean squared error loss, is then applied to capturing both global and local dependencies within single-omics data, extracting discriminative features for scRNA-seq and scATAC-seq. Subsequently, Inter-CL aligns these multi-omics features with cross-modal attention and learns cross-omics dependencies using multi-omics contrastive loss and multi-omics matching loss. The fused cellular representations can be applied in various downstream single-cell omics data analysis tasks.}\label{fig1}
\end{figure*}

We develop the scI$^2$CL framework to effectively fuse matched single-cell multi-omics data and learn discriminative cellular representations for various downstream tasks (Fig.~\ref{fig1}). scI$^2$CL consists of two major modules: \textit{intra-omics contrastive learning} (Intra-CL) and \textit{inter-omics contrastive learning} (Inter-CL). 
Given two single-cell omics datsets: one scRNA-seq omics datset and one scATAC-seq omics dataset, their count matrices are denoted by $X$ and $Y$, from which their multi-view masked matrices $X^m$ and $Y^m$ are constructed~(Step 1). First, Intra-CL extracts unified single-omics features using a zero-inflated negative binomial distribution-based autoencoder to obtain global and local features $Z_X, Z_Y$ and $Z_X^m, Z_Y^m$~(Step 2). As multihead attention is effective in local feature learning, Intra-CL uses it to generate unified local feature representations~(Step 3). To better unify the local and global information, Intra-CL employs the \textit{mean squared error} (MSE) loss as a soft fusion strategy for aligning the feature information while minimizing the interference of local noise with global information~(Step 4). Then, Inter-CL adopts cross-modal attention to preserve  the individuality of single-omics features $Z_r$ and $Z_a$ while capturing cross-omics dependencies~(Step 5). Here, the \textit{multi-omics contrastive}~(MOC) loss is used to maximize inter-omics similarity, thereby learning the specific gene-peak relationships between paired data~(Step 6). The aligned features $Z_{\text{r} \rightarrow \text{a}}$ and $Z_{\text{a} \rightarrow \text{r}}$ are eventually fused using a multi-layer perceptron~(Step 7). During this process, unpaired multi-omics data are combined to construct negative samples. The \textit{multi-omics matching}~(MOM) loss distinguishes between positive and negative samples, ensuring that the fusion focuses on the complementary omics information~(Step 8). The resulting integrated features $Z$ can be applied for various downstream analysis tasks, including cell clustering, cell subtype analysis, cell annotation correction, and cell development trajectory construction in this paper. The technical details of scI$^2$CL are presented in Section~\ref{sec:technical-details}.

\subsection{scI$^2$CL achieves outstanding performance in cell clustering}
\begin{figure*}
\centering
\includegraphics[width=0.9\textwidth]{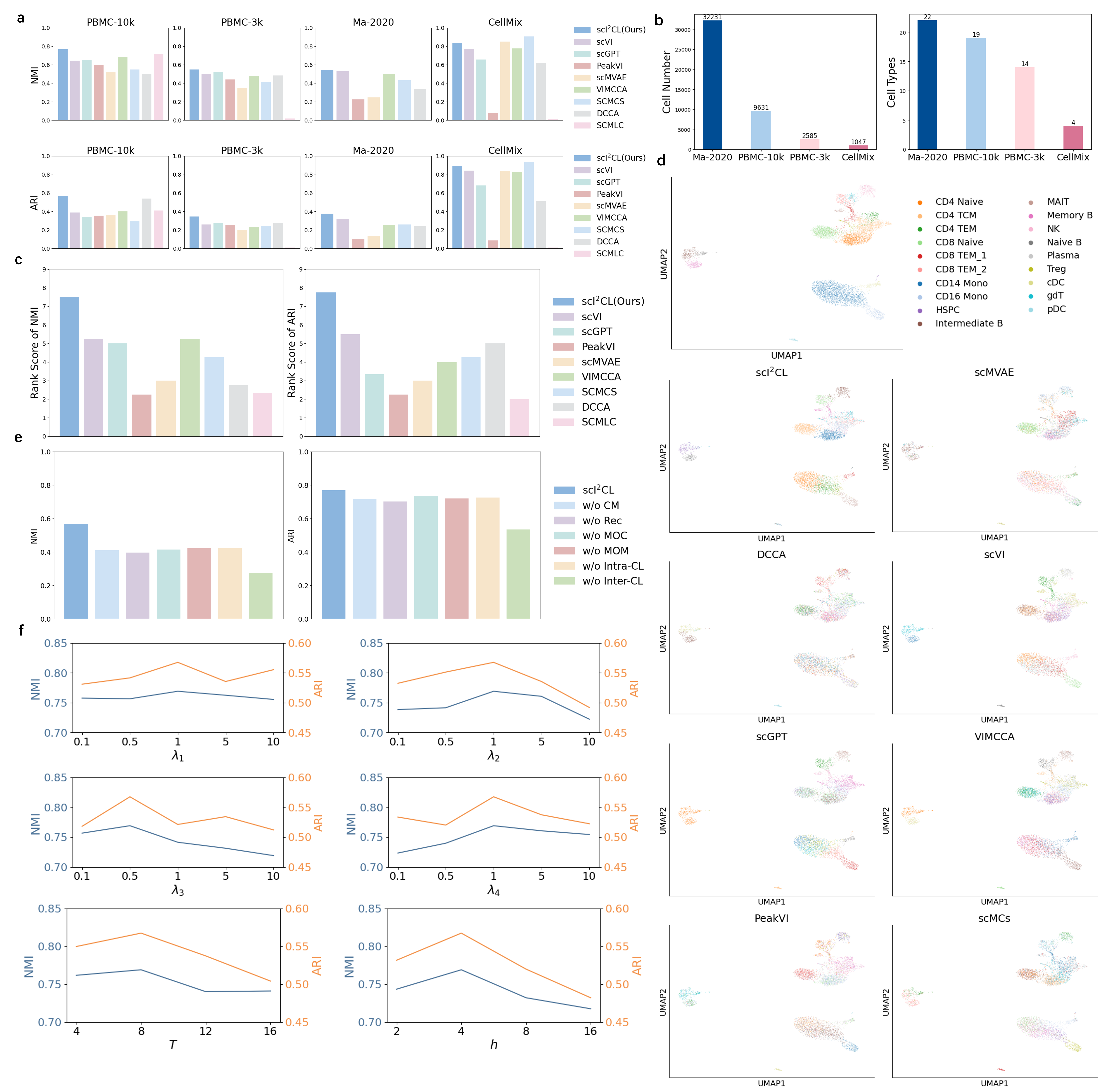}
\caption{Performance evaluation of scI$^2$CL on real-world single cell multi-omics datasets. \textbf{a} Performance comparison on four benchmark datasets using NMI and ARI. The x-axis denotes different methods, while the y-axis indicates the NMI/ARI score, where higher scores correspond to better clustering performance. \textbf{b} Statistical results of the numbers of cells (left panel) and cell types (right panel) in the four real-world datasets. \textbf{c} The $rs$ rank scores of different methods based on NMI and ARI across four datasets. Higher rank scores indicate better overall model performance.  \textbf{d} Visualization of the PBMC-10k dataset (human blood) and the corresponding clustering results by different  methods. \textbf{e} The results of ablation experiments on the PBMC-10k dataset, where different functional components were removed or replaced to demonstrate their impacts on the performance of scI$^2$CL. \textbf{f} Experimental results of the six hyperparameters' effects on the performance of scI$^2$CL.}\label{fig2}
\end{figure*}

We evaluate scI$^2$CL in cell clustering on four widely-used real world datasets: PBMC-10k, PBMC-3k, Ma-2020 and CellMix, and compare it with eight state-of-the-art  methods (SOTAs): SCMLC~\cite{scmlc}, DCCA~\cite{dcca}, scMCs~\cite{scmcs}, VIMCCA~\cite{vimcca}, scMVAE~\cite{scmvae}, PeakVI~\cite{peakvi}, scGPT~\cite{scgpt} and scVI~\cite{scvi}~(Details of these methods are in Section~\ref{sec:baselines}). Two evaluation metrics, Normalized Mutual Information~(NMI)~\cite{nmi} and Adjusted Rand Index~(ARI)~\cite{ari} are applied for assessing clustering accuracy. 
On three (PBMC-10k, PBMC-3k and Ma-2020) of the four datasets, scI$^2$CL outperform all SOTAs in terms of both NMI and ARI~(Fig.~\ref{fig2}a, Supplementary Table S1). Unlike the other datasets, CellMix contains only four cell types and 1,047 cells (Fig.~\ref{fig2}b, Supplementary Table S2), which limits the contrastive learning effect of scI$^2$CL and poses challenges for learning multi-omics relationships based on the discrepancies between different types of cells. Nevertheless, scI$^2$CL still ranks the second on CellMix. To further compare the overall performance of these methods, we introduce the rank score $rs$=$n$-$\bar{r}$~\cite{scmlc}, where $n$ is the total number of methods and $\bar{r}$ is the average rank of each method across all datasets. This metric showcases the global superiority of scI$^2$CL~(Fig.~\ref{fig2}c). Finally, UMAP visualizations of the PBMC-10k dataset and the clustering results of all methods on that dataset are illustrated to highlight scI$^2$CL’s advantage~(Fig.~\ref{fig2}d). Its intra-omics contrastive learning captures global and local dependencies in single-omics data, and inter-omics contrastive learning aligns multi-omics features, which leads to clear clustering boundaries and less cell types mixing compared to the other methods, demonstrating scI$^2$CL’s ability to learn discriminative cellular representations.

\begin{figure*}
\centering
\includegraphics[width=0.8\textwidth]{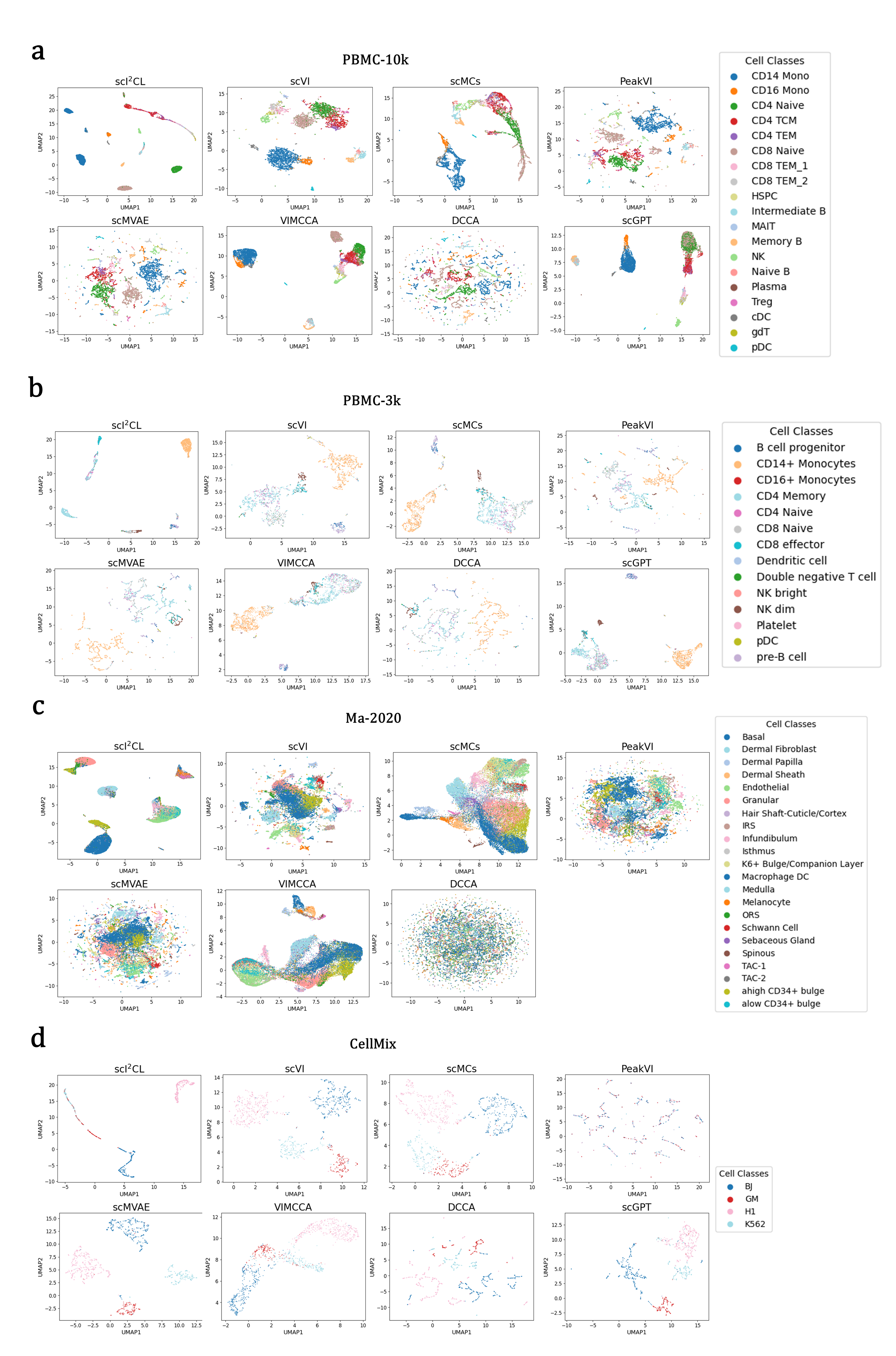}
\caption{The UMAP visualizations of cellular representations learnt by eight different methods across four real-world datasets. \textbf{a} PBMC-10K. \textbf{b} PBMC-3K. \textbf{c} Ma-2020. \textbf{d} CellMix. Note that on Ma-2020, there is no result for scGPT, because it is a human genes based pre-trained model, is ineffective for Ma-2020 --- a mouse cell dataset.}\label{fig3}
\end{figure*}

\subsection{Intra- and inter omics contrastive learning effectively improves cell clustering accuracy}
To verify the effectiveness of the proposed intra- and inter-omics contrastive learning, we conducted additional experimental validation and analysis.
On the PBMC-10K dataset, we evaluated the clustering performance of scI$^2$CL when various functional components were removed~(Fig.~\ref{fig2}e): 
1) \textbf{w/o CM}: the cross-modal attention for aligning multi-omics features is not utilized. 
2) \textbf{w/o Rec}: the zero-inflated negative binomial reconstruction loss for supervising single-omics feature learning is omitted.
3) \textbf{w/o MOC}: the multi-omics contrastive (MOC) loss for supervising multi-omics feature alignment is removed.
4) \textbf{w/o MOM}: the multi-omics matching (MOM) loss for supervising multi-omics feature integration is removed.
5) \textbf{w/o Intra-CL}: the intra-omics contrastive learning module that unifies the local-global relationships of single-omics features, is not employed. Instead, feature encoding relies solely on an autoencoder based on reconstruction loss.
6) \textbf{w/o Inter-CL}: the inter-omics contrastive learning module that preserves the individuality of single-omics features while capturing cross-omics dependencies, is not used. Integration of multi-omics features is performed exclusively via a multi-layer perceptron acting on concatenated features.
Results of the ablation study reveal that removing or replacing any component of scI$^2$CL leads to a significant decline in its capability of learning cellular representations. Notably, removing the crucial module of inter-CL prevents scI$^2$CL from effectively integrating omics individuality and specificity. This incurs a more substantial performance drop on real-world datasets than the removal of the reconstruction loss, which is vital in unsupervised learning scenes. Consequently, scI$^2$CL loses its performance advantage over the other methods.

Further hyperparameter experiments on the PBMC-10K dataset confirm the robustness of scI$^2$CL~(Fig.~\ref{fig2}f). 
In scI$^2$CL, $\lambda_1$, $\lambda_2$, $\lambda_3$ and $\lambda_4$ are used to balance the reconstruction loss, mean squared error loss, multi-omics contrastive loss, and multi-omics matching loss, respectively. $T$ is the number of multi-view matrices, and $h$ denotes the number of attention heads in the multi-head attention mechanism. To assess their impacts, we vary $\lambda$ value in~\{0.1,0.5,1,5,10\}, $T$ in~\{4, 8, 12, 16\} and $h$ in~\{2, 4, 8, 16\}. scI$^2$CL achieves the most competitive performance when $\lambda_1=1$, $\lambda_2=1$, $\lambda_3=0.5$ and $\lambda_4=1$. When $T$ is less than 8, scI$^2$CL exhibits degraded performance due to insufficient information. Conversely, as $T$ exceeds 8, redundant information increases, making it more challenging to train the model. For $h$, excessive heads produce excessively small feature segments, diminishing the effectiveness of the attention mechanism. Conversely, when $h$ is less than 4, the feature segments become too long, making it more difficult for the model to capture local dependencies. Despite hyperparameter effects, scI$^2$CL outperforms the compared methods even in worst-case scenarios, demonstrating its robustness. Additionally, the UMAP visualizations of cellular representations generated by different methods demonstrate that scI$^2$CL, leveraging its intra- and inter- contrastive learning, most distinctly captures the specificity of each cluster in two-dimensional space~(Fig.~\ref{fig3}).

\begin{figure*}
\centering
\includegraphics[width=0.8\textwidth]{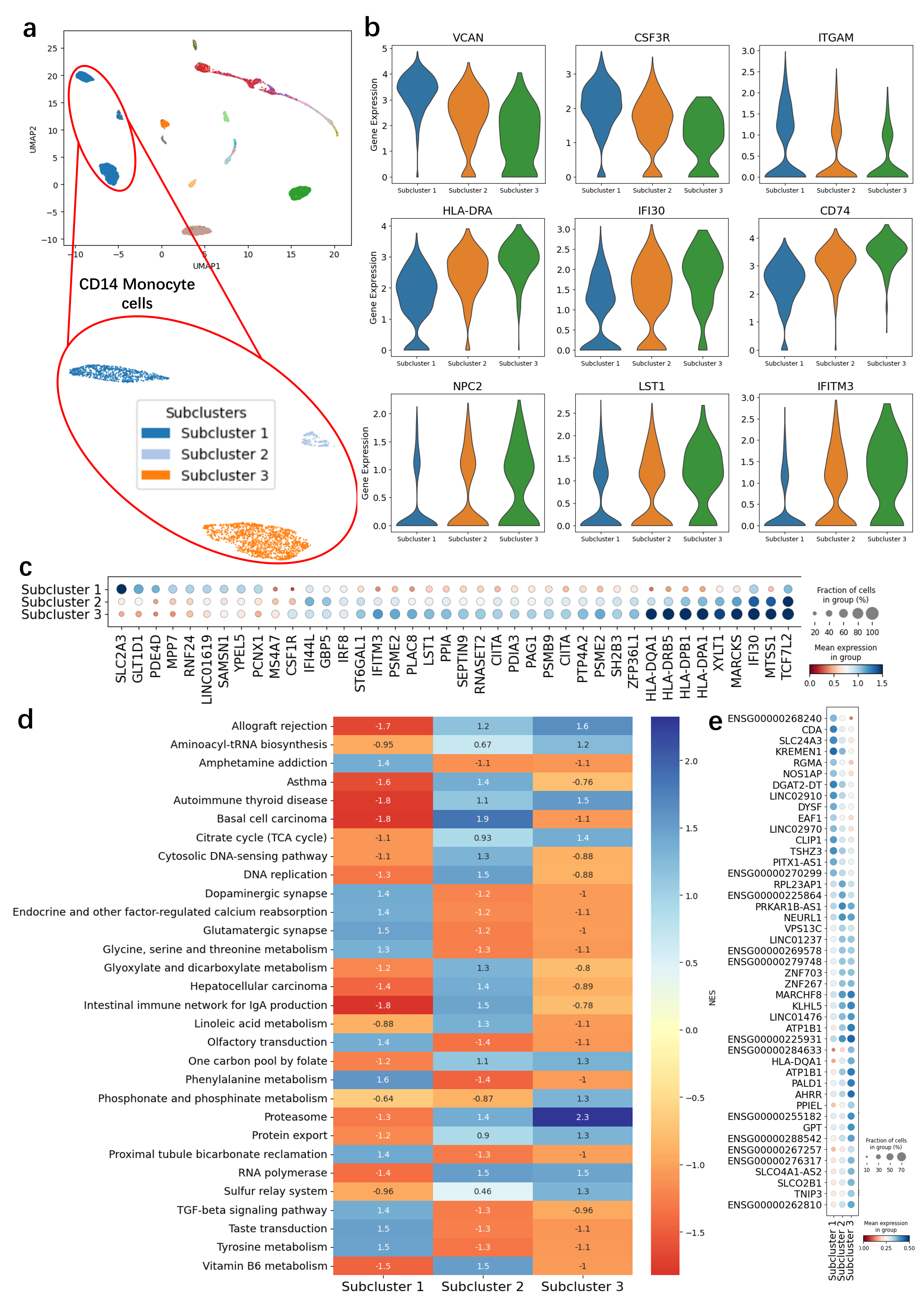}
\caption{scI$^2$CL reveals potential cellular subpopulations within CD14 Mono cells in PBMC-10K. \textbf{a} Visualization of cellular representations generated by scI$^2$CL reveals that the CD14 Monocytes form three distinct subclusters, which are potential cellular subpopulations --- not found by existing methods. \textbf{b} Violin plot visualization of partial gene markers in DEGs shows clear expression differences among the three subclusters. \textbf{c} Dot plot visualization of DEGs reveals substantial transcriptional differences among the three subclusters. \textbf{d} Heatmap visualization of pathway normalized enrichment scores from gene set enrichment analysis reveals evident cellular functional differences among the three subclusters. \textbf{e} Visualization of DEGs mapped from scATAC-seq data to the transcriptomic level shows significant heterogeneity across the three cell subclusters in both omics.}\label{fig4}
\end{figure*}

\subsection{scI$^2$CL identifies potential cellular subpopulations of CD14 Monocyte}
Our experiments on the PBMC-10K dataset further show scI$^2$CL's ability to uncover potential cell subpopulations, which cannot be found by the other methods. The visualization of cellular representations after dimensionality reduction derived from scI$^2$CL clearly distinguishes three subclusters within CD14 Mono, a type of monocyte characterized by high expression of the lipopolysaccharide (LPS) receptor CD14 and low expression of Fc$\gamma$IIIR CD16 (CD14++CD16-)~\cite{CD14_1}. These subpopulations are involved in pathogen recognition, antigen presentation, and immune regulation, and are associated with cardiovascular disease, cancer, and chronic inflammatory conditions~\cite{CD14_2}. To characterize the inherent heterogeneity of these cellular subpopulations, we resolved CD14 Monocytes into three distinct subclusters~(Fig.~\ref{fig4}a). To demonstrate the heterogeneity of the three subclusters, differentially expressed genes~(DEGs) are calculated using the Wilcoxon rank-sum test~\cite{wilcoxon} on normalized and preprocessed transcriptomic data for each subcluster, with a significance threshold of adjusted p-values being set to 0.05.
In the classic classification of human monocyte populations, DEGs encompass genes that serve as markers for distinct populations~\cite{CD14_3}. VCAN, CSF3R, ITGAM and S100A9  are markers for the classical population~(CD14++CD16-). HLA-DRA, IFI30, CD74, HLA-DPB1, SECTM1 and PPIA are associated with the intermediate population~(CD14++CD16+). NPC2, LST1, IFITM3, NAP1L1, CSF1R, MS4A7, TCF7L2, TAGLN, and MTSS1 are markers for the non-classical population~(CD14+CD16+)~\cite{CD14_marker1,CD14_marker2}. In a previous finer-granularity analysis of  four CD14 monocyte subsets associated with microscopic polyangiitis, DEGs such as S100A9, VCAN, IFI6, IFI44L, LY6E, ISG15, HLA-DRB1, and HLA-DMA are consistent with the highly expressed genes in these subsets~\cite{CD14_marker3}. 
Violin plots illustrate differences in gene markers among the three subclusters~(Fig.~\ref{fig4}b, Supplementary Figure S1). Furthermore, dot plots reveal global DEGs expression differences across the three subclusters, further supporting their classification into three distinct potential subpopulations~(Fig.~\ref{fig4}c). 
Further gene set enrichment analysis~(GSEA) on DEGs also highlights the differences among the three subclusters. Using the KEGG 2021 Human gene set as a reference and the prerank method for GSEA, a normalized enrichment score~(NES)-based heatmap highlights their potential functional differences in immunity, cancer, metabolize and other diseases or mechanisms~\cite{pathway1, pathway2, pathway3}~(Fig.~\ref{fig4}d).
We also conducted experiments on scATAC-seq data. 
We used EpiScanpy and GENCODE annotations to map the scATAC-seq data to the transcriptomic level~\cite{gencode, episcanpy} and calculated DEGs using the same strategy as before. Analysis on the scATAC-seq data also confirms the distinctiveness of the three subclusters~(Fig.~\ref{fig4}e).
Our abundant experimental results and prior studies collectively confirm that our proposed scI$^2$CL method successfully reveals latent CD14 Monocytes subpopulations. This finding underscores scI$^2$CL's robust feature representation capability benifitted from our intra-CL and inter-CL, enabling it to effectively capture cellular heterogeneity from single-cell multi-omics data in an unsupervised manner.

\subsection{scI$^2$CL rectifies mis-annotations of CD4+ Naive T cells as CD4+ Memory T cells}
\begin{figure*}
\centering
\includegraphics[width=0.8\textwidth]{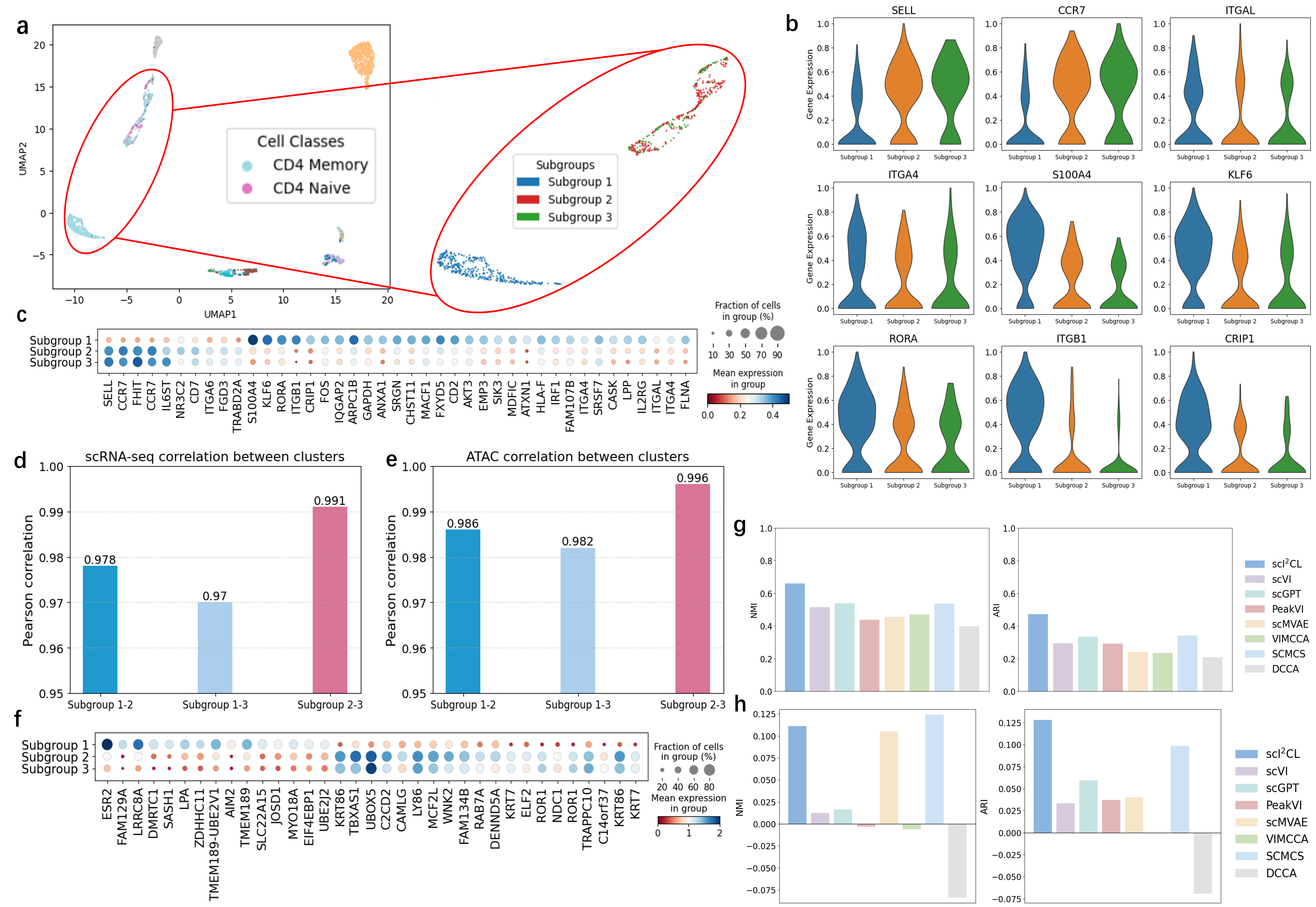}
\caption{scI$^2$CL corrects erroneous cell type labels for CD4+ Memory T  cells in PBMC-3k. \textbf{a} Visualization of cell representations reveals that CD4+ Memory T cells are partitioned into two distinct groups: one forms a clear cluster, while the other mixes with Naive T cells. \textbf{b} Violin plots visualizing DEGs demonstrate greater transcriptional similarity between Subgroup 2 and Subgroup 3 than to between Subgroup 2 and Subgroup 1. \textbf{c} Dot plot visualization of DEGs further confirms stronger global transcriptional correlation between Subgroups 2 and 3. \textbf{d} Pearson correlation coefficients of average transcriptome-wide expression shows the highest transcriptional similarity between Subgroups 2 and 3. \textbf{e} Pearson correlation coefficients of average gene activity scores derived from scATAC-seq data reveals the strongest similarity between Sub-groups 2 and 3. \textbf{f} Dot plot visualization of DEGs identified from scATAC-seq-derived gene activity scores confirms stronger similarity between Subgroups 2 and 3. \textbf{g} Clustering performance comparison among deep learning methods using the corrected labels. scI$^2$CL keeps superior performance. \textbf{h} Clustering performance improvement or degradation  achieved by deep learning methods with the corrected labels. }\label{fig5}
\end{figure*}

The experimental results on the PBMC-3K dataset demonstrate that scI$^2$CL can identify mislabeled cells. As shown in Fig.~\ref{fig5}a, 
the visualization of dimensional reduced cellular representations obtained from scI$^2$ CL reveals that CD4+ Memory T cells are distinctly separated into two subgroups. One of them forms a distinct cluster, we denote it as Subgroup 1 (blue), while the other group mixes with CD4+ Naive T cells and forms a mixed cluster, in which we denote the CD4+ Memory T cells as Subgroup 2 (red) and the CD4 + Naive T cells as Sub-Group 3~(green). The ontogenetic link and transitional differentiation states between CD4+ Naive T cells and their Memory counterparts, CD4+ Memory T cells, form a biological continuum, which complicates the classification of discrete cell types~\cite{CD4_1,CD4_2}. To facilitate further experimental analysis of their relationship patterns, we first performed differential gene expression analysis at the transcriptomic level using the Wilcoxon rank-sum test with an adjusted p-value threshold of 0.05. 
The analysis result shows that marker genes for CD4+ Naive T cells appear in DEGs, such as SELL (CD62L) and CCR7, which exhibited significantly higher expression in Subgroup 2 and 3. Conversely, Memory T cell marker genes ITGAL (CD11a) and ITGA4 (CD49d) showed significantly elevated expression in Subgroup 1 compared to the other subgroups~\cite{CD4_markgene1,CD4_markgene2,CD4_markgene3,CD4_markgene4}~(Fig.~\ref{fig5}b, Supplementary Figure S2). Furthermore, global visualization of the DEGs reveals that while cells in both Subgroup 1 and Subgroup 2 are annotated as CD4+ T Memory, cells in Subgroup 2 is more similar to that in Subgroup 3~(CD4+ T Naive cells) in transcriptomic level~(Fig.~\ref{fig5}c). This suggests a possible misannotation of cells in Subgroup 2 in the original PBMC-3k dataset. To verify the hypothesis, we calculated the average transcriptome-wide expression for all the subgroups and computed Pearson correlation coefficients between them. The results show significantly lower similarity between Subgroup 1 and Subgroup 2 than between Subgroup 2 and Subgroup 3, supporting the reclassification of cells in Subgroup 2 as CD4+ Naive T~(Fig.~\ref{fig5}d). For the scATAC-seq data from the PBMC-3k dataset, we utilized gene activity scores derived from peaks to compute Pearson correlation coefficients and identify DEGs. The analysis results on scATAC-seq also reveal that Subgroup 2 is more similar to Subgroup 3 than to Subgroup 1, providing additional support for cells in Subgroup 2 being misclassified as CD4+ Naive T~(Fig.~\ref{fig5}e, Fig.~\ref{fig5}f). To assess the sensitivity of different methods to this cell misclassification, we recalculate the Normalized Mutual Information (NMI) and Adjusted Rand Index (ARI) metrics for deep learning-based methods on the PBMC-3k dataset using the corrected labels. It turns out that scI$^2$CL maintains its overall superiority in terms of both metrics~(Fig.~\ref{fig5}g). Although most methods benefit from label correction, scI$^2$CL achieves the greatest improvement in ARI and the second improvement in NMI~(Fig.~\ref{fig5}h). In summary, scI$^2$CL can robustly capture cellular heterogeneity through Intra-CL and Inter-CL, which makes it able to clearly resolve these ambiguous cells.

\subsection{scI$^2$CL helps accurately construct developmental trajectories of human blood cells}
\begin{figure*}
\centering
\includegraphics[width=0.8\textwidth]{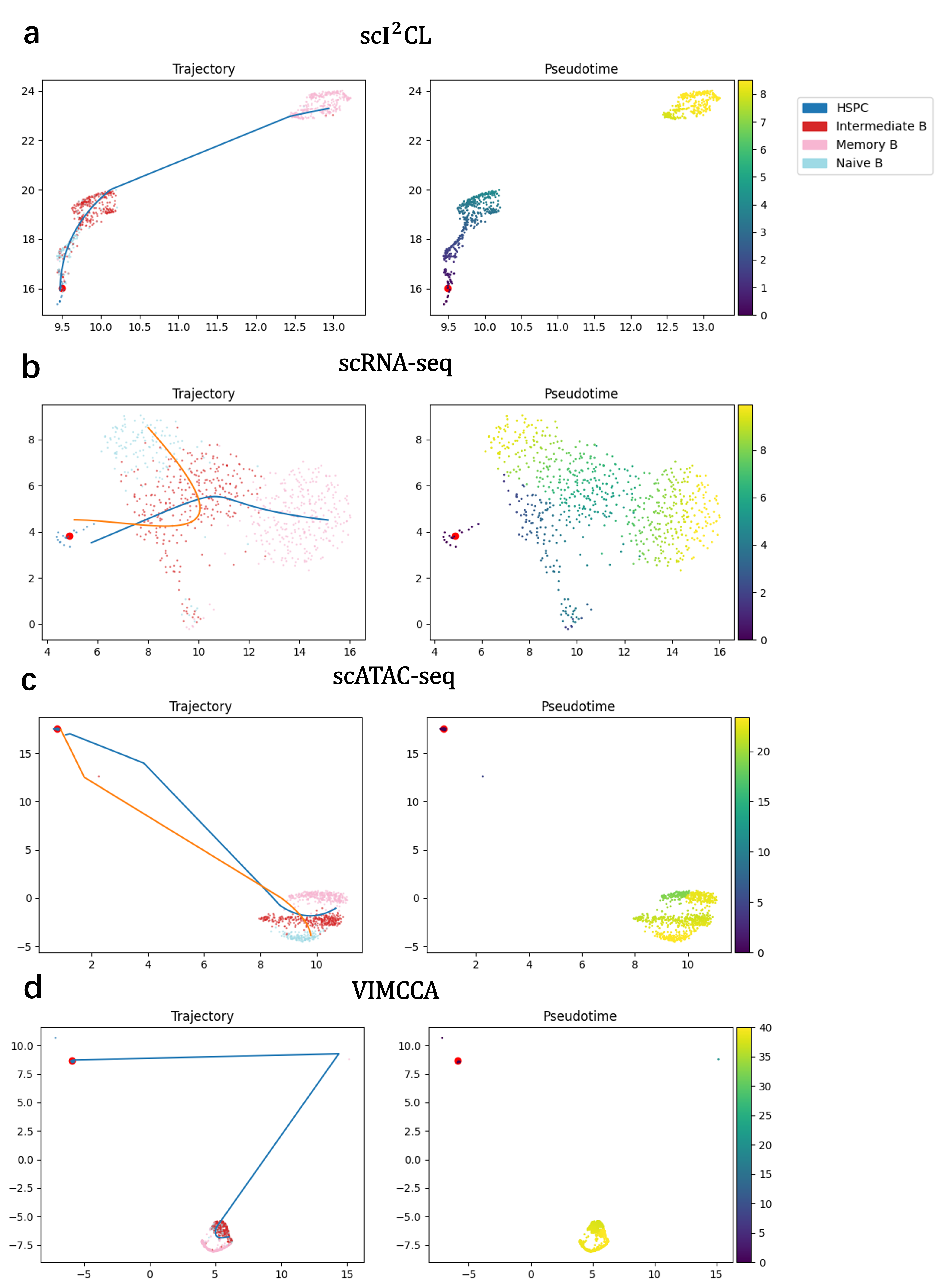}
\caption{The developmental trajectories and pseudo-time inferred from cellular representations generated by different data and methods. \textbf{a} The B cell developmental trajectory: HSPC $\rightarrow$ Naive B $\rightarrow$ Intermediate B $\rightarrow$ Memory B cells, constructed with cellular representations generated by scI$^2$CL. \textbf{b} Using only scRNA-seq data, two trajectories were inferred: one is a partially correct trajectory: HSPC $\rightarrow$ Intermediate B $\rightarrow$ Memory B cells, another is an erroneous trajectory: HSPC $\rightarrow$ Intermediate B $\rightarrow$ Naive B cells. \textbf{c} With only scATAC-seq data, two trajectories were also inferred: a partially correct trajectory: HSPC $\rightarrow$ Intermediate B $\rightarrow$ Memory B cells, and an erroneous trajectory: HSPC $\rightarrow$ Intermediate B $\rightarrow$ Naive B cells. \textbf{d} The developmental trajectory inferred with representations generated by VIMCCA: HSPC $\rightarrow$ Memory B $\rightarrow$ Naive B  $\rightarrow$ Intermediate B cells.}\label{fig6}
\end{figure*}

To further validate the efficacy of scI$^2$CL for integrating single-cell multi-omics data, we perform pseudo-time prediction and infer cell developmental trajectories using our cellular representations. With the PBMC-10K dataset, we first reduce our representations to two dimensions using UMAP, followed by pseudo-time inference and cell trajectory construction with  SlingShot~\cite{slingshot}, by designating hematopoietic stem and progenitor cells (HSPC), the primary source of human blood cell development, as the root node~\cite{hspc1}. The inference results reveal that the pseudo-time derived from the integrated representations learned by scI$^2$CL delineates a developmental path from HSPC, through Naive and Intermediate B cells, to Memory B cells~(Fig.~\ref{fig6}a, Supplementary Figure S3). Hematopoietic stem and progenitor cells, serving as the developmental origin of human blood cells, have been demonstrated to possess the capacity to differentiate into B lymphocytes~\cite{hspc2}. Moreover, the differentiation pathway from Naive to Intermediate to Memory B cells has been well-established through substantial previous studies~\cite{B_cell_1,B_cell_2,B_cell_3}. 
To check whether multi-omics fusion by scI$^2$CL does benefit accurate developmental trajectory construction, we compare the trajectory above (Fig.~\ref{fig6}a) with the inferred trajectories separately using scRNA-seq and scATAC-seq data. For scRNA-seq data, we applied Principal Component Analysis (PCA) to the raw gene count matrix, constructed a $k$-nearest neighbor (KNN) graph (n\_pcs = 30, n\_neighbors = 15), and performed dimensionality reduction using UMAP. For scATAC-seq data, we first reduced dimensionality via Latent Semantic Indexing (LSI) on the raw peak count matrix, constructed a KNN graph (n\_neighbors = 15), and then generated UMAP embeddings. 
The constructed trajectories from scRNA-seq and scATAC-seq data reveal two conserved paths: (1) HSPC $\rightarrow$ Intermediate B $\rightarrow$ Memory B (Fig.~\ref{fig6}b, Supplementary Figure S4), and (2) HSPC $\rightarrow$ Intermediate B $\rightarrow$ Naive B (Fig.~\ref{fig6}c, Supplementary Figure S5). These results demonstrate that scI$^2$CL's Intra-CL and Inter-CL can effectively integrate multi-omics data to yield biologically meaningful cellular representations, thus supporting trajectory inference better. Furthermore, we employ cellular representations generated by VIMCCA --- the second best method on the PBMC-10K dataset, to predict pseudo-time and construct the cellular developmental trajectory. This leads to a spurious HSPC $\rightarrow$ Memory B $\rightarrow$ Naive B cell $\rightarrow$ Intermediate B trajectory (Fig.~\ref{fig6}d, Supplementary Figure S6), further demonstrating the superiority of scI$^2$CL in single-cell multi-omics data integration.
Additionally, we computed the mean pseudo-time of the four types of cells by different methods (Supplementary Figure S7). Only multi-omics integration methods recapitulated the established developmental trajectory. Furthermore, compared to VIMCCA, the average pseudo-time inferred from the integrated representations generated by scI$^2$CL can be used to more accurately construct the cellular developmental trajectory, revealing the strong ability of scI$^2$CL for capturing cellular heterogeneity.

\section{Discussions}
In this study, we introduced scI$^2$CL, a novel deep learning method for learning unified representations that comprehensively capture cellular heterogeneity from single-cell multi-omics data. scI$^2$CL combines intra- and inter-omics contrastive learning for single-cell multi-omics integration. On the one hand, it performs intra-omics contrastive learning within individual omics to derive high-quality feature representations by aligning local and global information. On the other hand, it employs inter-omics contrastive learning to uncover potential cross-omics relationships between different omics. Specifically, scI$^2$CL first generates multiple distinct sub-views via random masking, then employs multi-head attention to integrate feature embeddings from these sub-views, utilizing an MSE-dependent soft fusion strategy that minimizes the noise from redundant information while combining global and local context features in a single omics. Furthermore, scI$^2$CL employs cross-modal attention to enable feature interaction between different omics for preserving omics-specific information. To prevent inherent data noise (e.g., sequencing deviations and batch effects) from obscuring critical cross-omics interactions, the multi-omics contrastive loss maximizes the mutual information between paired omics data. Meanwhile, scI$^2$CL constructs positive and negative sample pairs from matching/unmatching single-cell multi-omics data and applies the multi-omics matching loss, which makes the model focus on learning the cross-omics relationships from the truly matching cells, yielding integrated representations that comprehensively model cellular heterogeneity. In general, scI$^2$CL is an unsupervised learning method applicable for various downstream tasks. By using only gene-peak count matrices, it robustly captures cellular heterogeneity representations, making scI$^2$CL highly flexible for various single-cell multi-omics datasets without needing external information, enabling a range of downstream analyses including cell clustering, cell subpopulation analysis, cell annotation correction and cell development trajectory construction.

Compared with eight leading domain-specific methods, scI$^2$CL consistently demonstrated superior performance across four real-world datasets in cell clustering. Dimensionality reduction visualizations of cellular representations further reveal that scI$^2$CL produces clear clustering boundaries. Furthermore, scI$^2$CL identified three latent subpopulations within human monocytes --- a discovery unachievable by the other computational tools, exhibiting their significant differences in transcriptomic and epigenomic expression and function. Simultaneously, scI$^2$CL resolved cell type misclassification of CD4+ Memory T cells, and distinguishing correctly labeled and mislabeled memory T cells into distinct subgroups. Additionlly, only when using the cellular representations learned by scI$^2$CL, we correctly constructed the HSPC → Naive B → Intermediate B → Memory B cell developmental trajectory. These extensive experimental results demonstrate that scI$^2$CL can effectively integrate multi-omics data, yields discriminative cellular representations, thus accurately elucidates cellular heterogeneity.

While scI$^2$CL has demonstrated effective integration of single cell multi-omics sequencing data, significant opportunities for improvement still remain. First, although its intra- and inter-omics contrastive learning captures heterogeneity across cells, modeling nuanced state transitions (e.g., cellular senescence or pathogenic transformation) within the same cell types requires further investigation.  Second, as scI$^2$CL relies on matched multi-omics data to construct contrastive samples, how to extend this paradigm to unpaired multi-omics data poses a promising yet challenging future research direction. 

In conclusion, scI$^2$CL is an effective computational method for intergating single-cell multi-omics data, by leveraging the intra- and inter-omics contrastive learning strategy, it can generate comprehensive and discriminative cellular representations that are aware of cellular heterogeneity, thus can be applied in precisely elucidating and characterizing cross-omics patterns among cells.

\section{Methods}
\subsection{Data preprocessing}
scI$^2$CL is trained and infers using only single-cell count matrices, enabling its flexible application to the integration of various single-cell multi-omics data.
In this paper, we use four typical single-cell joint profiling datasets, including 10x-Multiome-PBMC 10k dataset~(PBMC-10K), 10x-Multiome-PBMC 3k dataset~(PBMC-3K), mouse skin cells dataset~(Ma-2020)~\cite{ma2020} and human cell line mixture dataset~(CellMix)~\cite{chen2019}.
PBMC-10K comprises 9,631 human peripheral blood mononuclear cells across 19 distinct cell types. The Ma-2020 dataset contains 32,231 murine skin cells spanning four batches, falling into 22 cell types. Dimensionality reduction visualization reveals no substantial batch effects across the four batches in the Ma-2020 dataset~(Supplementary Figure S8). Consequently, we preprocessed and utilized the dataset as a unified entity for model training.
PBMC-3K contains 2,585 human peripheral blood mononuclear cells across 14 distinct cell types. And the CellMix dataset contains 1047 cells of four cell types: BJ, H1, K562 and GM12878.
Detailed statistics for these four real-world datasets are given in Supplementary Table S2.

\subsection{Design of scI$^2$CL}\label{sec:technical-details}
\subsubsection{Intra-omic contrastive learning for single-omics feature extraction} 
Given the scRNA-seq count matrix $X$$\in$$R ^ {N \times D_X}$ and the scATAC-seq count matrix $Y$$\in$$R ^ {N \times D_Y}$, where $N$ is the number of cells, $D_X$ and $D_Y$ are the number of genes and peaks, respectively. scI$^2$CL first generates multi-view matrices $X^{m}$ and $Y^{m}$ by randomly masking gene or peak columns and uses two independent autoencoders $f_x$, $f_y$ to learn the local feature $Z_{X}$, $Z_{Y}$ and the global feature $Z_{X}^{m}$, $Z_{Y}^{m}$:
\begin{equation}\label{encoder}
\begin{aligned}
Z_{X} = f_x(X,\theta_X)&,\quad Z_{Y} = f_y(Y,\theta_Y)\\
Z_{X}^{m} = f_x(X^m,\theta_X)&,\quad Z_{Y^m} = f_y(Y^m,\theta_Y)
\end{aligned}
\end{equation}
where \{$Z_{X}$,$Z_{Y}$\} $\in$$R ^ {N \times d}$ and \{$Z_{X}^{m}$,$Z_{Y}^{m}$\} $\in$$R ^ {T \times N \times d}$, $d$ is the dimension of embedding space, $T$ is the number of multi-view matrices, $\theta_X$ and $\theta_Y$ are learnable parameters of the encoders.
To extract better single-omics features, we propose \textit{intra-omics contrastive learning} (Intra-CL) to capture both global and local dependencies within the single-omics data. Concretely, scI$^2$CL integrates multi-view features with multi-head attention~\cite{attention} followed by max pooling as follows:
\vspace{-0.2cm}
\begin{equation}\label{mhatt}
\begin{aligned}
&\hat{Z}_{\{X,Y\}} = 
   \text{maxpooling}\left(
   \text{mulhead}\left(Z^m_{\{X,Y\}}\right)
   W^O\right)\\
   &\text{mulhead}
   \left(Z^m_{\{X,Y\}}\right)=\text{head}_1\left(Z^m_{\{X,Y\}}\right)
   \oplus \cdots \oplus 
   \text{head}_h\left(Z^m_{\{X,Y\}}\right) 
   \\
&\text{head}_i\left(Z^m_{\{X,Y\}}\right) =
   \text{softmax}\left(
   \frac{Q K^\top}{\sqrt{d_k}}
   \right) V \\
&Q = Z^m_{\{X,Y\}} W_{Q_i}, \quad
 K = Z^m_{\{X,Y\}} W_{K_i}, \quad
 V = Z^m_{\{X,Y\}} W_{V_i}
\end{aligned}
\end{equation}
where $W_{Q_i}$$\in$$ R^{d \times d_k}$, $W_{K_i}$$\in$$ R^{d \times d_k}$, $W_{V_i}$$\in$$ R^{d \times d_k}$, and $W^O$$\in$$ R^{d \times d}$ are learnable parameters, $\oplus$ is the concatenating operation, $h$ is the number of heads, $d_k$$=$$d/h$ is head dimension, and \{$\hat{Z}_X$, $\hat{Z}_Y$\} $\in$$R^{N \times d}$ are integrated multi-view features. Multi-head attention plays a crucial role in encoding single-omics features. The attention mechanism allows multi-view features to interact with one another, while the multi-head structure enables the model to learn from multiple feature segments across different perspectives. This enables the model to effectively captures local dependencies within the features. 
Following this, max pooling is applied to downsampling the multi-view features, which also enhances prominent local features to preserve the captured critical local dependencies.

To capture global dependencies in single-omics data, we employ two independent decoders $g_x$ and $g_y$ to reconstruct the single-omics data using $Z_X$ and $Z_Y$.
Considering the discrete and highly sparse nature of single-cell data, we utilize the zero-inflated negative binomial~(ZINB) distribution as the reconstruction loss~\cite{zinb}. ZINB incorporates the mean $\mu$ and dispersion $\theta$ parameters from the negative binomial distribution, along with a coefficient $\pi$ that describes the probability of dropout events as follows:
\begin{equation}\label{zinb1}
\begin{aligned}
NB(x; \mu, \theta) = \frac{\Gamma(x + \theta)}{\Gamma(\theta)} \left( \frac{\theta}{\theta + \mu} \right)^\theta \left( \frac{\mu}{\theta + \mu} \right)^x\\
ZINB(x; \pi, \mu, \theta) = \pi \zeta_0(x) + (1 - \pi) NB(x; \mu, \theta)
\end{aligned}
\end{equation}
where $x$ is a vector from the single-omics data \{$X,Y$\}, $\Gamma$ denotes the gamma function, and $\zeta_0(x)$ is the probability distribution of zero values in $x$. 

Concretely, the decoders estimate parameters of the ZINB distribution by three separate fully connected layers as follows:
\begin{equation}\label{zinb2}
\begin{aligned}
\Pi_{\{x, y\}} &= sigmoid(g_{\{x,y\}}(Z_{\{X,Y\}}, \phi^{\pi}_{\{X,Y\}}))\\
\overline{M}_{\{x, y\}}&=\exp(g_{\{x,y\}}(Z_{\{X,Y\}}, \phi^{\mu}_{\{X,Y\}}))\\
\Theta_{\{x, y\}} &= \exp(g_{\{x,y\}}(Z_{\{X,Y\}}, \phi^{\theta}_{\{X,Y\}}))
\end{aligned}
\end{equation}
where $\Pi_{\{x, y\}}$,$\overline{M}_{\{x, y\}}$ and $\Theta_{\{x, y\}}$ represent the matrix forms of $\pi$, $\mu$ and $\theta$ for both omics, $ \phi^{\pi}_{\{X,Y\}}$, $\phi^{\mu}_{\{X,Y\}}$ and $\phi^{\theta}_{\{X,Y\}}$ are learnable parameters of three fully connected layers for $g_x$ and $g_y$, respectively. Since the dropout probability $\pi$ falls in [0, 1], we use the $sigmoid$ activation function for computing coefficient. For $\overline{M}$ and $\Theta$, we apply the exponential function $exp$ as the activation function, considering the non-negative nature of mean and dispersion parameters.
The reconstruction loss for the decoder network is defined as the negative logarithm of the ZINB likelihood, i.e.,
\begin{equation}\label{zinb3}
\begin{aligned}
\mathcal{L}_{rec}=-log(ZINB(X|\Pi_x,\overline{M}_x,\Theta_x)) -log(ZINB(Y|\Pi_y,\overline{M}_y,\Theta_y))
\end{aligned}
\end{equation}

To simultaneously capture both local and global dependencies in single-omics data, we introduce a soft fusion strategy using mean squared error~(MSE) loss to integrate single-modal features with multi-view features, thereby facilitating intra-omics contrastive learning. Formally, 
\begin{equation}\label{mse}
\begin{aligned}
\mathcal{L}_{MSE}=\frac{1}{N} \sum_{i=1}^{N} (Z_i - \hat{Z}_i)^2
\end{aligned}
\end{equation}
where $Z$ $\in$ $\{Z_X, Z_Y\}$ denotes the single-modal features, $\hat{Z}$ $\in$ $\{\hat{Z}_X,\hat{Z}_Y\} $ denotes the multi-view features.

\subsubsection{Inter-omic contrastive learning for multi-omics feature integration} 
Although intra-omics contrastive learning improves the feature representations of single-omics data, it also makes the model pay more attention to the individuality of each omics and the features more vulnerable to noise, which poses challenges for the integration of omics features.

To address the challenges, we propose \textit{inter-omics contrastive learning} (Inter-CL) to guide the model in capturing inter-omics dependencies and filtering noise, thereby enhancing cell representations. Concretely, scI$^2$CL first aligns multi-omics features with cross-modal attention, a variant attention used to capture latent cross modal adaptation~\cite{mult}. The cross-modal attention from modality scRNA-seq~($r$) to modality scATAC-seq~($a$) can be formulated as
\begin{equation}\label{cm_att1}
\begin{aligned}
\text{CM}_{\text{$r$} \rightarrow \text{$a$}}(Z_r,Z_a) = \text{softmax}\left(
    \frac{rW_{Q_r}W^\top_{K_{a}}a^\top}{\sqrt{d}}
    \right)aW_{V_{a}}\\ 
\end{aligned}
\end{equation}
where ${Z_r, Z_a}\in R^{N \times d} $ denote the feature vector of modality $r$ and $a$ respectively, $W_{Q_r}$, $W_{K_a}$ and $W_{V_a}$ are learnable parameters. 
To align the single-omics features with each other, we use bidirectional cross-modal attention $\text{CM}_{\text{r} \rightarrow \text{a}}$ and $\text{CM}_{\text{a} \rightarrow \text{r}}$, that is,  
\begin{equation}\label{cm_att2}
\begin{aligned}
Z_{\text{r} \rightarrow \text{a}}=\text{CM}_{\text{r} \rightarrow \text{a}}(Z_X,Z_Y), \quad Z_{\text{a} \rightarrow \text{r}}=\text{CM}_{\text{a} \rightarrow \text{r}}(Z_Y,Z_X) 
\end{aligned}
\end{equation}
where $Z_{\text{r} \rightarrow \text{a}}$ denotes the aligned scRNA-seq features and $Z_{\text{a} \rightarrow \text{r}}$ denotes the aligned scATAC-seq features. 
Benefiting from the cross-modal attention, single-omics features preserve their individuality while capturing dependencies across different omics, leading to improved representation of cellular states.

Additionally, we employ \textit{multi-omics contrastive}~(MOC) loss to supervise cross-omics interactions. The MOC loss facilitates the commonality between paired data while filtering the noise in unmatched ones, thereby enabling the model to more effectively capture inter-omics dependencies and allowing the features to provide more accurate representations of cells~\cite{albef}. Concretely, given a batch of $n$ cells, we firstly compute softmax-normalized similarity by
\begin{equation}\label{moc1}
\begin{aligned}
p^{r2a}_{i,j}= \frac{exp(s_{i,j})}{\sum_{k=1}^n exp(s_{i,k})} \quad
p^{a2r}_{i,j}= \frac{exp(s_{i,j})}{\sum_{k=1}^n exp(s_{k,j})}
\end{aligned}
\end{equation}
where $s_{i,j}, \{i,j\}\in[0,n] $ denotes the similarity of the $i$-th scRNA-seq feature and the $j$-th scATAC-seq feature, \{$p^{r2a}$, $p^{a2r}$\} $\in R^{n \times n}$ are scRNA-seq to scATAC-seq and scATAC-seq to scRNA-seq similarities.
Let $y^{r2a}$ and $y^{a2r}$ denote the ground-truth one-hot similarities, where paired ones have a probability of 1 and unpaired ones have a probability of 0. The  multi-omics contrastive loss is defined as the cross-entropy between $p$ and $y$. Formally,
\begin{equation}\label{moc2}
\begin{aligned}
\mathcal{L}_{MOC}=-\frac{1}{2}(\sum_iy^{r2a}logp^{r2a} + \sum_iy^{a2r}logp^{a2r}) 
\end{aligned}
\end{equation}

Then, we concatenate the aligned single-omics features $Z_{\text{r} \rightarrow \text{a}}$ and $Z_{\text{a} \rightarrow \text{r}}$, and apply a multi-layer perceptron for feature integration:

\begin{equation}\label{mlp}
\begin{aligned}
Z=f_M(Z_{\text{r} \rightarrow \text{a}}\textcircled{+}Z_{\text{a} \rightarrow \text{r}},W_Z)
\end{aligned}
\end{equation}
where $f_M$ is the multi-layer perceptron and $W_Z$ are learnable parameters. Additionally, we generate negative samples $Z_{neg}$ by concatenating and integrating unmatched single-omics features while using $Z$ as the positive examples. A \textit{multi-omics matching}~(MOM) loss is applied to distinguishing between $Z$ and $Z_{neg}$: 
\begin{equation}\label{mom}
\begin{aligned}
\hat{y}^Z=f_C(Z\textcircled{+} Z_{neg},W_C)\\
\mathcal{L}_{MOM}=-\sum_i y^Z_i log\hat{y}^Z_i
\end{aligned}
\end{equation}
where $f_C$ is a fully connected layer that projects the integrated features into two-dimensional vectors, $W_C$ indicates learnable parameters, and $y^Z$ indicates the one-hot labels of the positive and negative samples. By discriminating between positive and negative pairs, the model is able to capture intricate inter-omics associations.

\subsubsection{Loss function of scI$^2$CL} Combining the components above, the overall loss function of the scI$^2$CL model is defined as follows:
\begin{equation}\label{loss}
\begin{aligned}
\mathcal{L}=\lambda_1\mathcal{L}_{rec}+\lambda_2\mathcal{L}_{MSE}+\lambda_3\mathcal{L}_{MOC}+\lambda_4\mathcal{L}_{MOM}
\end{aligned}
\end{equation}
where $\lambda_1$, $\lambda_2$, $\lambda_3$ and $\lambda_4$ are scalar hyperparameters that weight $\mathcal{L}_{rec}$, $\mathcal{L}_{MSE}$, $\mathcal{L}_{MOC}$ and $\mathcal{L}_{MOM}$, respectively. By optimizing Equ.~(\ref{loss}), scI$^2$CL can effectively capture intra-omics and inter-omics dependencies within single-cell multi-omics data and extract integrated cell representations for downstream tasks, including cell clustering, cell subpopulation analysis, cell annotation correction, and cell development trajectory construction in this paper.

\subsection{Implementation details}
We employ the SCANPY~\cite{scanpy} pipeline for single-cell data preprocessing, including filtering low-quality data, selecting highly variable genes, and performing normalization to mitigate batch effects. scI$^2$CL is implemented under the pytorch deep learning framework and trained with a NVIDIA GeForce RTX 3090 GPU. The hyper-parameters $\lambda_1$, $\lambda_2$, $\lambda_3$, $\lambda_4$ ,$T$ and $h$ are set to 1, 1, 0.5, 1, 8 and 4, respectively. The training procedure of scI$^2$CL includes two steps. First, we pretrain the autoencoder-decoder networks for 200 epoches by minimizing the ZINB reconstruction loss. Then we introduce the intra and inter-omics contrastive learning, and train the scI$^2$CL model for 400 epoches with the reconstruction loss $\mathcal{L}_{rec}$, the mean squared error loss $\mathcal{L}_{MSE}$,  multi-omics contrastive loss $\mathcal{L}_{MOC}$ and multi-omics matching loss $\mathcal{L}_{MOM}$. We use an Adam optimizer with the learning rate 1e-5 to optimize our model, the batch size of each dataset is set to 128, and the masking percentage for multi-view masked matrices $X^m$/$Y^m$ is set to 60\%. 

For cell clustering, we apply $k$-means on the integrated representations. To facilitate comparison across methods, the value of $k$ is set to the number of annotated cell types in each dataset. 
For cell subpopulation analysis, we perform dimensionality reduction using UMAP~\cite{umap} and visualize the resulting embeddings. We utilize SCANPY's built-in methods for identification and visualization of differentially expressed genes (DEGs). Gene Set Enrichment Analysis (GSEA) is implemented using its Python package~\cite{gseapy}. The integration of scATAC-seq with scRNA-seq data is performed with EpiScanpy, using gene annotations retrieved from the GENCODE website.
For cell annotation correction, we employed the same methodology as used in cell subpopulation analysis to identify and visualize DEGs. Pearson correlation coefficients between subgroups were computed with SciPy's implementation~\cite{scipy}.
For cell development trajectory construction, we infer the cellular developmental trajectories using a Python implementation of slingshot, based on the UMAP embeddings generated by each method.

\subsection{Baselines}\label{sec:baselines}
We compare our proposed scI$^2$CL model with 8 state-of-the-art methods, including SCMLC~\cite{scmlc}, DCCA~\cite{dcca}, scMCs~\cite{scmcs}, VIMCCA~\cite{vimcca}, scMVAE~\cite{scmvae}, PeakVI~\cite{peakvi}, scGPT~\cite{scgpt} and scVI~\cite{scvi}.
\begin{itemize}
\item \textbf{SCMLC}~\cite{scmlc} captures modality-specific and cross-modality consistency information by constructing multiple single-modality and cross-modality cell-cell networks and employs a robust multimodal community detection approach to identify reliable cell clusters with multi-omics data.

\item \textbf{DCCA}~\cite{dcca} utilizes variational autoencoders to model data from individual omics, while leveraging an attention transfer mechanism to facilitate information interaction between different omics.

\item \textbf{scMCs}~\cite{scmcs} employs attention mechanisms and contrastive learning to capture both the individuality and commonality within single-cell multi-omics data and introduces a novel multi-clustering approach to identify cell states from multiple perspectives.

\item \textbf{VIMCCA}~\cite{vimcca} uses shared latent variables to capture commonalities across different omics data, while jointly learning two omics-specific non-linear models with the variational inference.

\item \textbf{scMVAE}~\cite{scmvae} proposes a probabilistic Gaussian mixture model based variational autoencoder and integrates multi-omics data with three strategies.

\item \textbf{PeakVI}~\cite{peakvi} models an informative latent space that preserves biological heterogeneity while correcting batch effects and accounting for technical effects. Additionally, it offers a technique for identifying differential accessibility at single-region resolution, enabling effective analysis for scATAC-seq data.

\item \textbf{scGPT}~\cite{scgpt} is a transformer based model pretrained over 33 million human cells, which can effectively extract critical biological insights concerning genes and cells from scRNA-seq data.

\item \textbf{scVI}~\cite{scvi} aggregates information from similar cells and genes through stochastic optimization based network while approximating the distribution of observed expression values, thus achieving scRNA-seq clustering analysis.
\end{itemize}

We use Normalized Mutual Information~(NMI)~\cite{nmi} and Adjusted Rand Index~(ARI)~\cite{ari} to evaluate clustering performance. The value of NMI falls in $[0,1]$ and ARI falls in $[-1,1]$, where higher values indicate better clustering performance. NMI is  employed to evaluate the consistency between predicted and ground truth labels, which is defined as follows:
\begin{equation}
\text{NMI}(T,Y)=\frac{2MI(T,Y)}{H(T)+H(Y)}
\end{equation}
where $T=\{t_1, t_2,\cdots, t_n\}$ and $Y=\{y_1, y_2, \cdots, y_n\}$ represent the ground truth and predicted labels. $MI(T, Y)$ denotes mutual information between $T$ and $Y$, while $H$ represents the information entropy.

ARI is used to measure the similarity between predicted and ground truth labels, which is defined as follows:
\begin{equation}\label{ARI}
\text{ARI} = \frac{\sum_{ij} \binom{n_{ij}}{2} - \frac{\sum_i \binom{a_i}{2} \sum_j \binom{b_j}{2}}{\binom{n}{2}}}
{\frac{1}{2} \left[ \sum_i \binom{a_i}{2} + \sum_j \binom{b_j}{2} \right] - \frac{\sum_i \binom{a_i}{2} \sum_j \binom{b_j}{2}}{\binom{n}{2}}}
\end{equation}
where $n$ denotes the number of cells in the dataset, $n_{ij}$ denotes the number of cell samples in the intersection of the ground truth cluster $C_i$ and the predicted cluster $K_j$. $a_i$ and $b_j$ represent the number of samples in the ground truth cluster $C_i$ and the predicted cluster $K_j$. $\binom{n}{2}$ denotes the binomial coefficient, i.e., $\frac{n\left(n-1\right)}{2}$.

\section*{Data availability}
All datasets used in this paper are freely available from the public source. The processed h5ad files of PBMC-10K and Ma-2020 datasets can be downloaded from scglue~(\href{https://scglue.readthedocs.io/zh-cn/latest/data.html}{https://scglue.readthedocs.io/zh-cn/latest/data.html}). The processed h5 file of PBMC-3K dataset can be downloaded from scMDC~(\href{https://github.com/xianglin226/scMDC/tree/master/datasets}{https://github.com/xianglin226/scMDC/tree/master/datasets}). 
And the tsv count matrices files of the CellMix dataset can be downloaded from GEO database~(\href{https://www.ncbi.nlm.nih.gov/geo/query/acc.cgi?acc=GSE126074}{GSE126074}).

\section*{Declarations}

\begin{itemize}

\item This work was supported by National Natural Science Foundation of China (NSFC) under grants No.~62172300, No.~62372326 and No.~62372116.

\item Author contribution: Conceptualization: WC.L, JH.G and SG.Z. Methodology: WC.L. Software: WC.L. Formal analysis: WC.L and H.P. Investigation: WC.L. Visualization: WC.L. Validation: WC.L. Writing and revision: WC.L, H.P, WG.L, YC.Z, SG.Z and JH.G. Supervision: JH.G and SG.Z.

\item The authors declare no competing interests.

\item Code availability: The source code of this work is available at \href{https://github.com/PhenoixYANG/scICL}{https://github.com/PhenoixYANG/scICL}

\end{itemize}

\selectlanguage{english}
\FloatBarrier
\phantomsection
\bibliography{ref} 

\end{CJK}\end{document}


\begin{CJK}{UTF8}{gbsn}

\title{scI$^2$CL: Effectively Integrating Single-cell Multi-omics by Intra- and Inter-omics Contrastive Learning}

\makeatletter
\renewcommand{\thefigure}{S\arabic{figure}}  
\renewcommand{\fnum@figure}{Supplementary Figure~S\arabic{figure}}  
\makeatother
\setcounter{figure}{0}
\makeatletter
\renewcommand{\thetable}{S\arabic{table}}  
\renewcommand{\fnum@table}{Supplementary Table~S\arabic{table}}  
\makeatother
\setcounter{table}{0}

\author[1]{Wuchao Liu}%
\author[1]{Han Peng}%
\author[1]{Wengen Li}%
\author[1]{Yichao Zhang}%
\author[1,$\ast$]{Jihong Guan}%
\author[2,$\ast$]{Shuigeng Zhou}%

\affil[1]{School of Computer Science and Technology, Tongji University, Shanghai 201804, China}%
\affil[2]{College of Computer Science and Artificial Intelligence, Fudan University, Shanghai 200438, China}%
\affil[*]{Corresponding author. \href{email:email1}{jhguan@tongji.edu.cn}
and \href{email:email2}{sgzhou@fudan.edu.cn}}

\vspace{-1em}

\date{ }

\begingroup
\let\center\flushleft
\let\endcenter\endflushleft
\maketitle
\endgroup

\selectlanguage{english}

\begin{figure}
\centering
\includegraphics[width=0.8\textwidth]{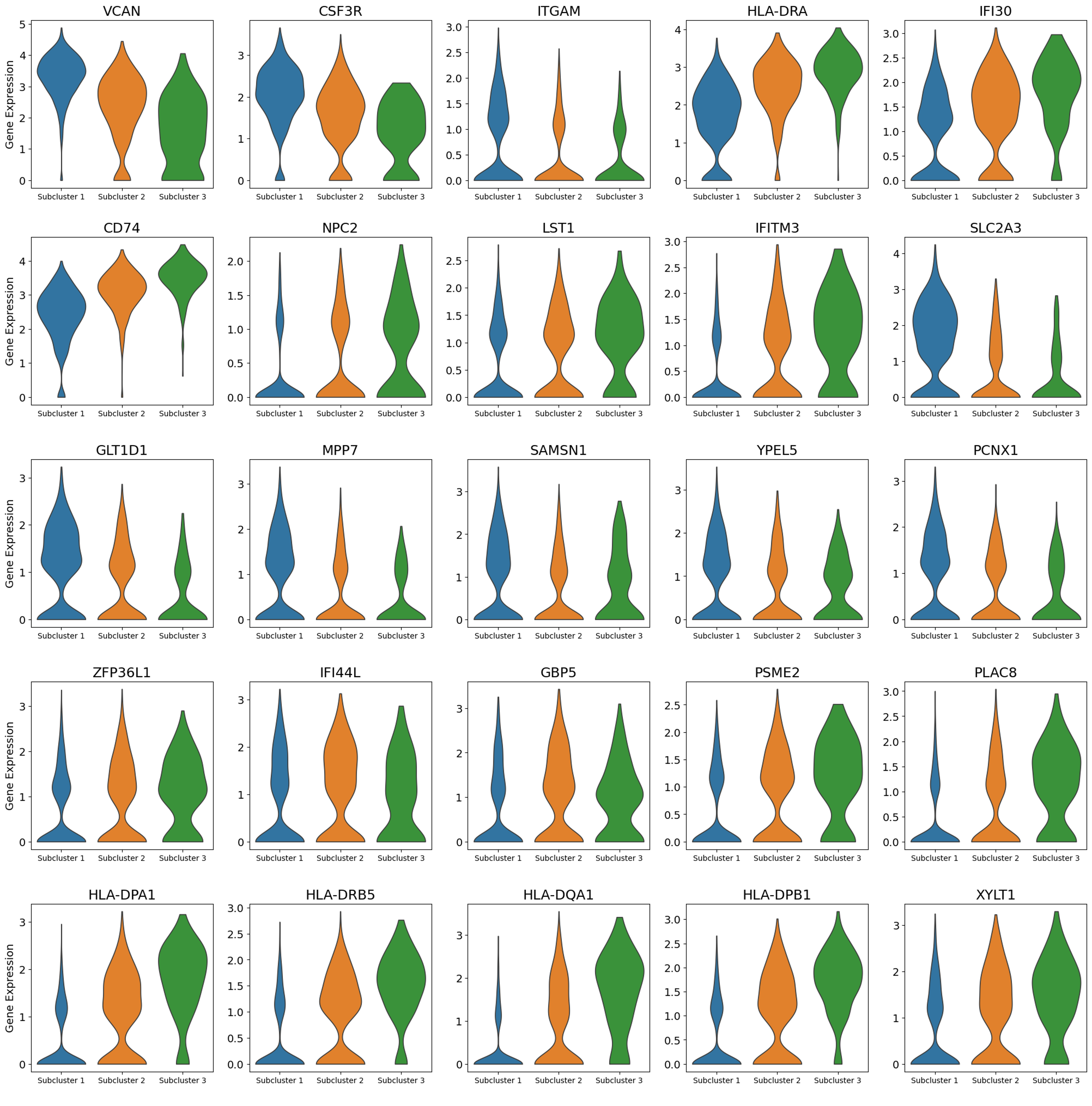}
\caption{Violin plots of  differentially expressed genes in three different CD14 Monocyte cellular subpopulations revealed by scI$^2$CL.  }
\label{scICL_slignshot}
\end{figure}

\begin{figure}
\centering
\includegraphics[width=0.8\textwidth]{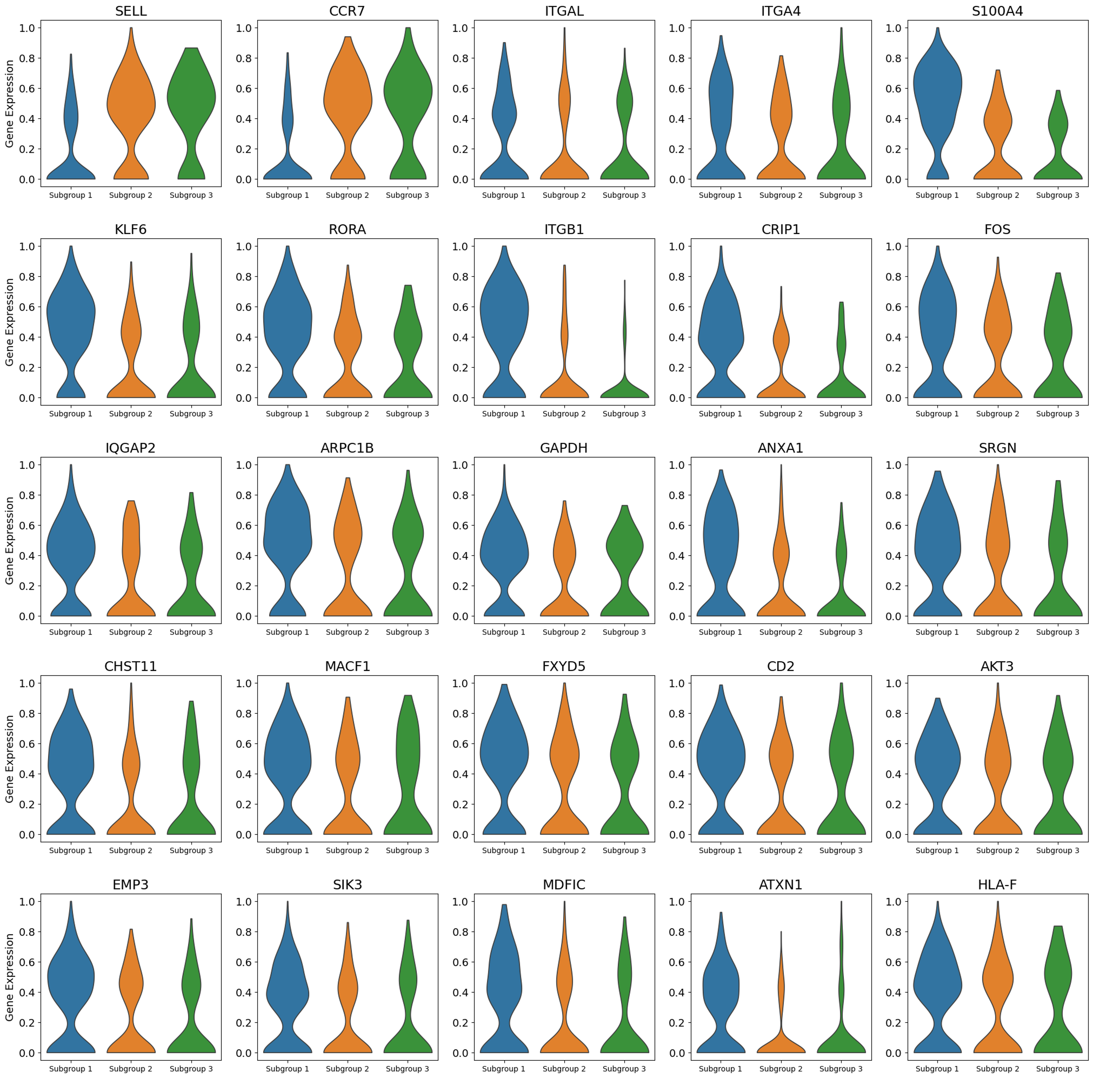}
\caption{Violin plots of  differentially expressed genes in three CD4+ T cell sub-groups defined by scI$^2$CL.  }
\label{scICL_slignshot}
\end{figure}

\begin{figure}
\centering
\includegraphics[width=0.8\textwidth]{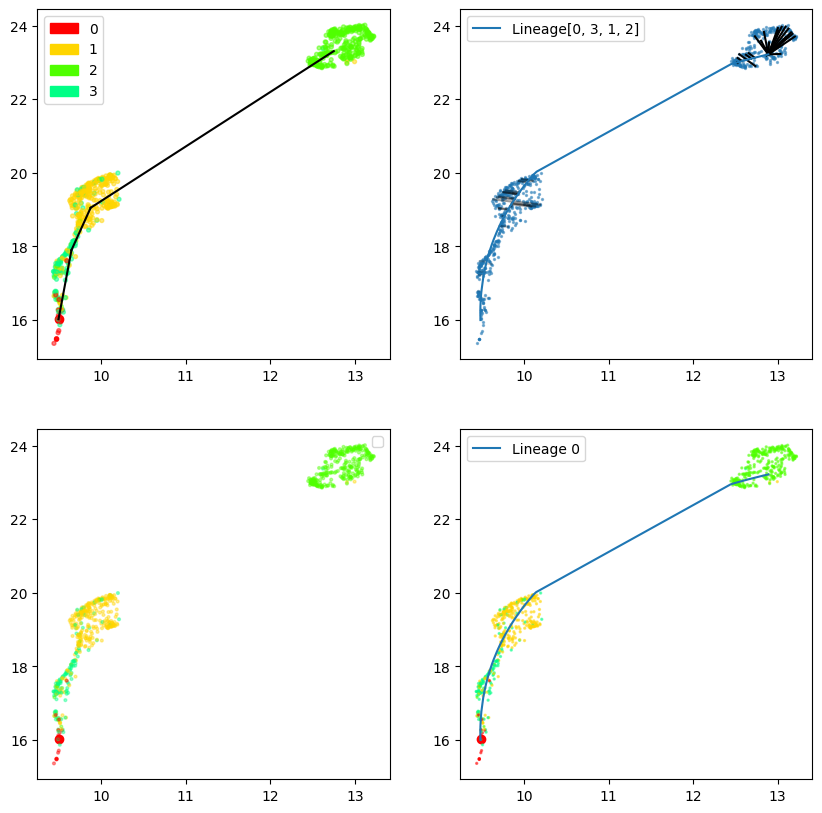}
\caption{Cell trajectory predicted with UMAP representations of scI$^2$CL.  }
\label{scICL_slignshot}
\end{figure}

\begin{figure}
\centering
\includegraphics[width=0.8\textwidth]{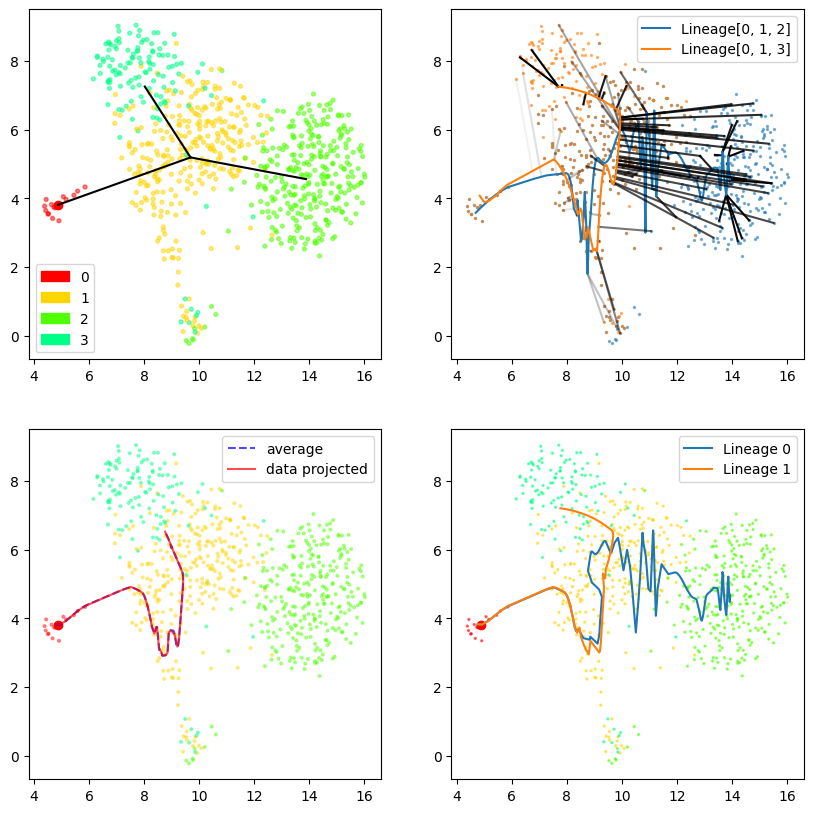}
\caption{Cell trajectory predicted with UMAP representations of scRNA-seq.  }
\label{scRNA_slignshot}
\end{figure}

\begin{figure}
\centering
\includegraphics[width=0.8\textwidth]{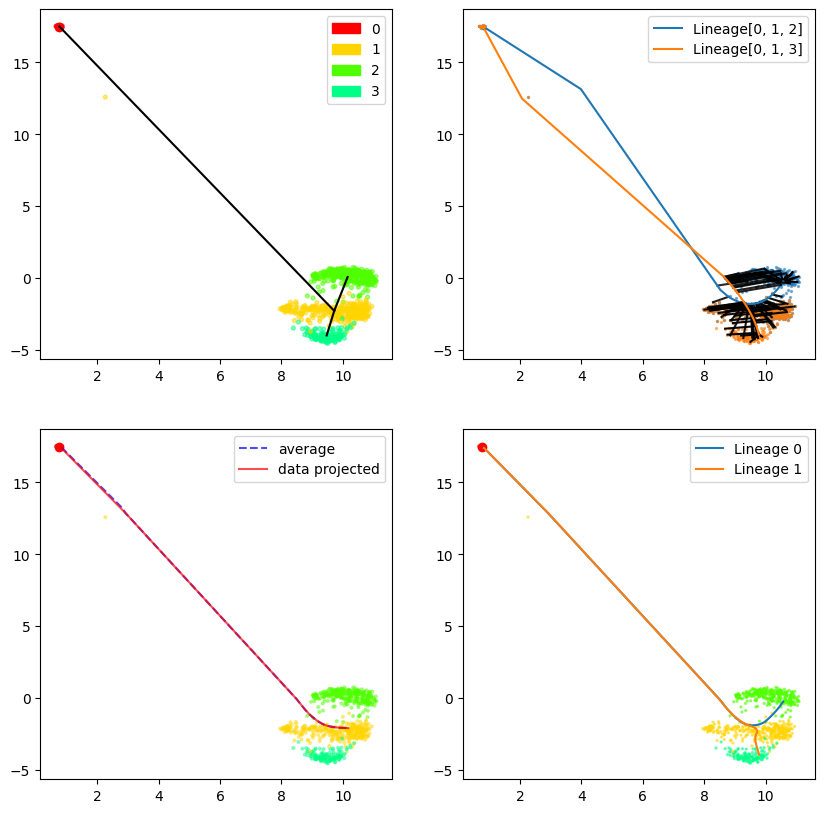}
\caption{Cell trajectory predicted with UMAP representations of scATAC-seq.  }
\label{scATAC_slignshot}
\end{figure}

\begin{figure}
\centering
\includegraphics[width=0.8\textwidth]{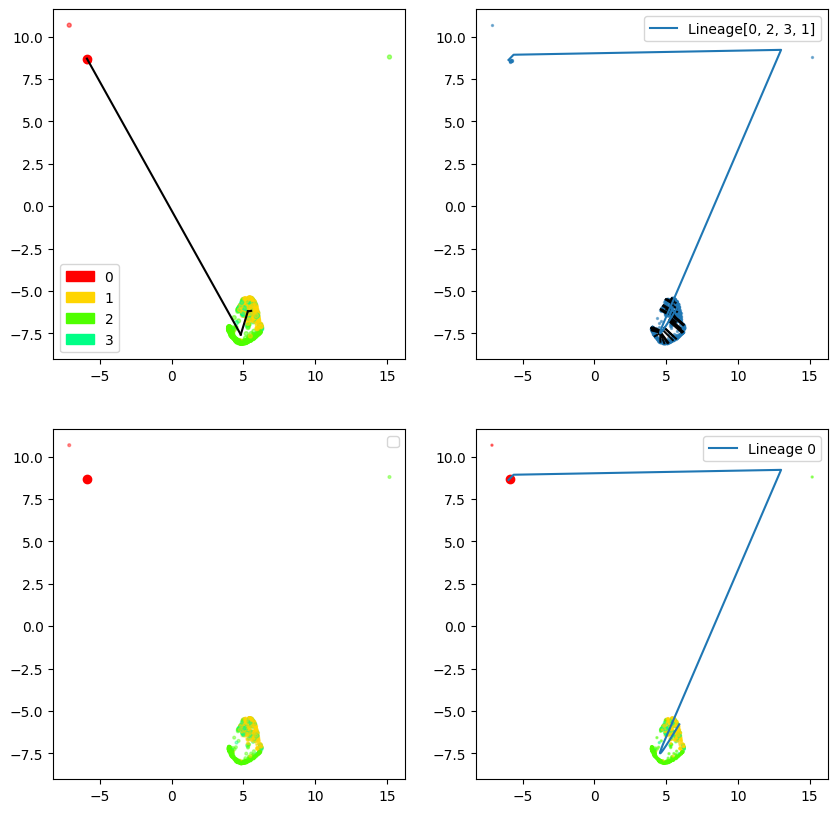}
\caption{Cell trajectory predicted with UMAP representations of scATAC-seq.  }
\label{VIMCCA_slignshot}
\end{figure}

\begin{figure}
\centering
\includegraphics[width=0.8\textwidth]{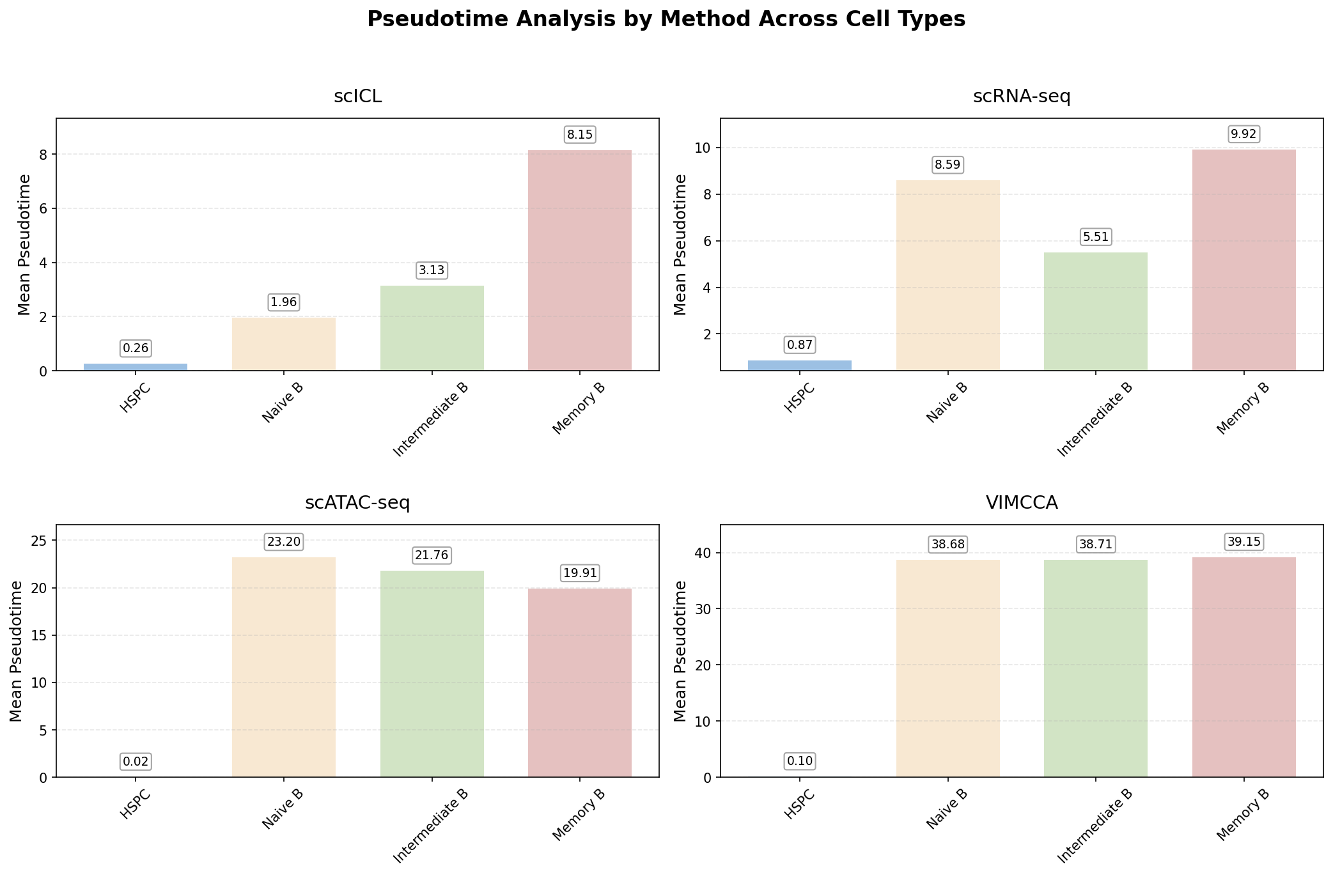}
\caption{Mean pseudo-time  visualization for four cell types across different methods.}
\label{Pseudotime}
\end{figure}

\begin{figure}
\centering
\includegraphics[width=0.8\textwidth]{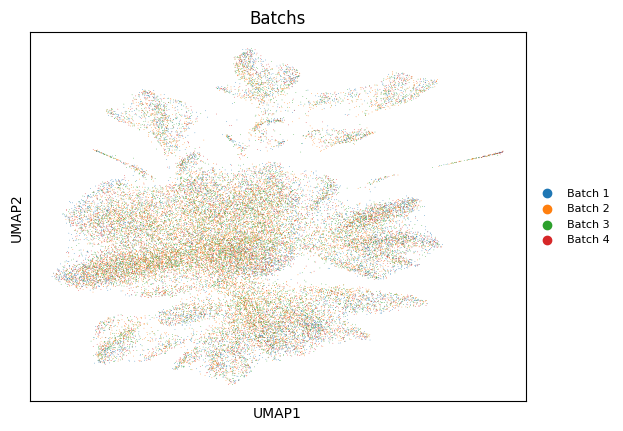}
\caption{Visualization of batch effects in the Ma-2020 dataset revealed no substantial differences across batches.}
\label{ma_batch}
\end{figure}

\begin{table}[h]
\caption{Comparison result of different methods on four real world single-cell multi-omics datasets.}
\label{compare}
\centering
\begin{tabular}{ccccccccc}
    \hline \hline
    \multicolumn{1}{c}{\multirow{2}{*}{Method}} & \multicolumn{2}{c}{PBMC-10K} & \multicolumn{2}{c}{PBMC-3K} & \multicolumn{2}{c}{Ma-2020} & \multicolumn{2}{c}{CellMix} \\
    \cline{2-9} & NMI & ARI & NMI & ARI & NMI & ARI & NMI & ARI  \\
\cmidrule(r){1-9}
scI$^2$CL~(Ours) &0.769 &0.567&0.549&0.344&0.543&0.376&0.835&0.893\\
scVI &0.644 &0.387&0.503&0.260&0.530&0.320&0.771&0.841\\
scGPT &0.650 &0.339&0.523&0.274&$\backslash$&$\backslash$&0.657&0.683\\
PeakVI&0.599 &0.356&0.440&0.254&0.225&0.102&0.077&0.085\\
scMVAE&0.519 &0.362&0.351&0.200&0.246&0.135&0.852&0.839\\
VIMCCA&0.687 &0.401&0.477&0.235&0.504&0.251&0.778&0.822\\
SCMCS&0.550 &0.294&0.412&0.243&0.433&0.260&0.907&0.939\\
DCCA&0.500 &0.540&0.483&0.278&0.337&0.240&0.619&0.513\\
SCMLC&0.718 &0.411&0.018&0.010&$\backslash$&$\backslash$&0.010&0.010\\
\hline \hline
\end{tabular}
\end{table}

\begin{table}
\caption{Statistics of four real world single-cell multi-omics datasets, where Genes and Peaks denote the dimension of scRNA-seq and scATAC-seq count matrix, respectively.}\label{dataset}
\centering
\begin{tabular}{l l l l c}
\hline \hline
Dataset &   Cells &  Genes &  Peaks &  Cell types\\
\midrule
PBMC-10K &  9631 & 29095 & 107194 &19\\
PBMC-3K &  2585 & 36601 & 20010 &14\\
Ma-2020 &  32231 & 21478 & 340341 &22\\
CellMix &  1047 & 18666 & 136771 &4\\
    \hline \hline
\end{tabular}
\end{table}

\end{CJK}